\documentstyle[12pt]{article}
\begin{document}
\newcommand{\fourj}[4]{\left[ \begin{array}{cc}
                              #1 & #2 \\
							  #3 & #4
							  \end{array} \right]}
\newcommand{\sixj}[7]{\left[ \begin{array}{cc}
                              #1 & #2 \\
							  #3 & #4 \\
							  ~~~~~~~ #5    \\
							  #6 & #7
							  \end{array} \right]}
\newcommand{\eightj}[8]{\left( \begin{array}{cccc}
                              #1 & #2 & #3 & #4 \\
							  #5 & #6 & #7 & #8
							  \end{array} \right)}
\rightline {IMSc./93/3}
\rightline{ Feb. 15, 1993}
\vspace{.4cm}

\begin{center}

{\bf Chern-Simons Theory, Coloured-Oriented Braids and Link Invariants}

\vspace{1cm}

{\bf R. K. KAUL}$^\star$ \\
{\it The Institute of Mathematical Sciences} \\
{\it C.I.T.Campus, Taramani,} \\
{\it Madras \ 600 113, India.}

\end{center}

\vspace{1cm}

\baselineskip=24pt

\noindent {\bf Abstract}

A method to obtain explicit and complete topological solution of SU(2)
Chern-Simons theory on $S^3$ is developed. To this effect the necessary aspects
of the theory of coloured-oriented braids and duality properties of conformal
blocks for the correlators of $SU(2)_k$ Wess-Zumino conformal field theory are
presented. A large class of representations of the generators of the groupoid
of coloured-oriented braids are obtained. These provide a whole lot of new link
invariants of which Jones polynomials are the simplest examples. These new
invariants are explicitly calculated as illustrations for knots upto eight
crossings and two-component multicoloured links upto seven crossings.

\vspace{2.5cm}

\hrule
\vspace{.5cm}
{\footnotesize $^\star$ email : kaul@imsc.ernet.in}

\newpage

\noindent {\bf 1. Introduction}

\vspace{.7cm}

Topological quantum field theories provide a bridge between quantum physics on
one hand and geometry and topology of low dimensional manifolds on the
other$^1$. The functional integral formulation of such quantum field theories
provides a framework to study this relationship. In particular, a class of
topological field theories which are related to knot theory have attracted a
good deal of attention in recent times. This started with the seminal work of
Witten who not only put the Jones polynomials$^2$ in a field theoretic setting,
but also presented a general field theoretic framework in which knot theory
could be studied in an arbitrary three-manifold$^3$.

In $SU(2)$ Chern-Simons gauge theory, the expectation value of Wilson link
operators with doublet representation placed on all the component knots yields
Jones polynomials. Two variable generalization of these polynomials, so called
HOMFLY polynomials$^4$, are obtained as the expectation value of Wilson link
operators with $N$ dimensional representation on all the component knots in an
$SU(N)$ Chern-Simons theory. In fact Witten$^3$ has shown that the expectation
values of such link operators obey the same Alexander-Conway skein relation as
those by Jones and HOMFLY polynomials respectively. These relations can be
recursively solved to obtain these polynomials for an arbitrary link. Placing
arbitrary representations on the component knots, corresponding generalizations
of Alexander-Conway relations can also be obtained$^{5,6}$. But unfortunately
these relations can not be solved recursively to obtain the link invariants.
Therefore there is a need to develope methods which would allow direct
calcuations of expectation values of Wilson link operators with arbitrary
representations living on the component knots. In refs.6, an attempt was made
to develope one such method. This allowed us to obtain invariants for links
that can be constructed from braids made of upto four strands. However, links
related to braids with larger number of strands still stayed elusive. Another
interesting method based on the construction of knot operators has also been
developed$^{7}$. It allows readily calculation of invariants for torus knots in
an elegant way. However, the scope of this method also appears to be limited
and it cannot be applied to obtain invariants for other knots.

In this paper we shall present a general and simple method of obtaining the
expectation value for an arbitrary Wilson link operator in an $SU(2)$
Chern-Simons gauge theory on $S^3$. The method can be generalised to other
compact non-abelian gauge groups as well as to manifolds others than $S^3$.

The $SU(2)$ Chern-Simons action is given by :
$$
k S \ = \ \frac{k}{4\pi} \int_{S^3} tr (AdA + \frac{2}{3} A^3) \eqno(1.1)
$$

\noindent where $A$ is a matrix valued connection one-form of the gauge group
$SU(2)$. The topological operators of this topological field theory are given
in terms of Wilson loop (knot) operators :
$$
W_j [C] \ = \ tr_j P exp \oint_C A \eqno(1.2)
$$

\noindent for an oriented knot $C$ carrying spin $j$ representation. These
operators are independent of the metric of the three-manifold. For a link $L$
made up of oriented component knots $C_1, C_2, \ldots C_s$ carrying spin $j_1,
j_2,\ldots j_s$ representations respectively, we have the Wilson link operator
defined as
$$
W_{j_1j_2\ldots j_s} [L] \ = \ \prod_{\ell=1}^s \ W_{j_\ell} [C_\ell]
\eqno(1.3)
$$

\noindent We are interested in the functional averages of these operators :
$$
V_{j_1j_2\ldots j_s}[L] \ = \ Z^{-1} \int_{S^3} [dA] W_{j_1j_2\ldots j_s} [L]
e^{ikS} \eqno(1.4)
$$
$$
Z \ = \ \int_{S^3} [dA] e^{ikS} \hspace{1.3cm}  \eqno(1.5)
$$

\noindent Here both the integrand in the functional integrals as well as the
measure are metric independent$^8$. Therefore, these expectation values depend
only on the isotopy type of the oriented link $L$ and the set of
representations $j_1, j_2\ldots j_s$ associated with the component knots. The
partition function above is given by$^3$ :
$$
Z \ = \ \sqrt{2/(k+2)}~~ sin (\pi / (k+2)).
$$

The method of calculating the functional averages (1.4) developed in the
present paper makes use of two important ingredients :

(i) The first ingredient is the intimate relationship that Chern-Simons theory
on a three-manifold with boundary has with corresponding Wess-Zumino conformal
field theory on that boundary$^{3,9,7}$. Consider a 3-manifold with a number of
two dimensional boundaries $\sum^{(1)}, \sum^{(2)}, \ldots \sum^{(r)}$. Each of
these boundaries, $\sum^{(i)}$ may have a number of Wilson lines carrying spins
$j_\ell^{(i)}, \ell =1,2,\ldots.$ ending or beginning at some points
(punctures) $P^{(i)}_\ell$ on them. Following Witten$^{3}$, we associate with
each boundary $\sum^{(i)}$ a Hilbert space ${\cal{H}}^{(i)}$. The Chern-Simons
functional integral over such a three-manifold is then given as a state in the
tensor product of these Hilbert spaces, $\otimes^r~~ {\cal{H}}^{(i)}$. The
operator formalism developed in ref.7, gives an explicit representation of
these states
as well as determines the form of the inner products of vectors belonging to
these Hilbert spaces. The conformal blocks of the $SU(2)_k$ Wess-Zumino field
theory on these boundaries $\sum^{(i)}$ with punctures $P_\ell^{(i)},
\ell=1,2,\ldots $ carrying spins $j^{(i)}_\ell$ determine the properties of
the associated Hilbert spaces ${\cal{H}}^{(i)}$. In fact, these provide a basis
for these Hilbert spaces ${\cal{H}}^{(i)}$. There are more than one possible
basis. These different bases are related by duality of the correlators of the
Wess-Zumino conformal field theory. We shall need to write down these duality
matrices explicitly for our discussion here.

(ii) The second input we shall need is the close connection knots and links
have
with braids. Theory of braids, first developed by Artin, is generally studied
for identical strands$^{10,11}$. What we need for our purpose here is instead a
theory of coloured and oriented braids. The individual strands are coloured by
putting
different $SU(2)$ spins on them. The necessary aspects of the theory of such
braids will be developed here. In particular a theorem due to Birman$^{11}$
relating links to plats of braids will be restated for coloured-oriented
braids. This theorem along with the duality properties of conformal blocks of
correlators in $SU(2)_k$ Wess-Zumino conformal field theory on an $S^2$ then
will allow us to present an explicit and complete solution of $SU(2)$
Chern-Simons gauge theory on $S^3$. Alternatively, a theorem due to Alexander
relating closure of braids to links can also be stated for coloured-oriented
braids. This theorem also provides an equivalent method of solving Chern-Simons
gauge
theory.

The knot invariants have also been extensively studied from the point of view
of exactly
solvable models in statistical mechanics$^{12,14}$. Wadati {\it et al}  have
exploited the intimate connection between exactly solvable lattice models with
knot invariants to obtain a general method for constructing such invariant
polynomials$^{14}$. Besides these, knot invariants have also been studied from
the point of view of quantum groups$^{15}$.

This paper is organised as follows. In section 2, we shall write down the
duality matrices relating two convenient complete sets of conformal blocks for
the $SU(2)_k$ Wess-Zumino conformal field theory on an $S^2$ with
$2m~~ (m=2,3,\ldots)$ punctures carrying arbitrary $SU(2)$ spins. Next, in
section
3, the required aspects of the theory of coloured-oriented braids will be
developed. A theorem due to Birman$^{11}$ relating oriented links with plats
will be
restated for plats of coloured-oriented braids. Alexander's theorem$^{16}$
relating closure of braids with oriented links can also be restated for
coloured-oriented links. In section 4, a class of representations of the
generators of braid groupoid will be presented. These ingredients then will
allows us to write down the complete and explicit solution of $SU(2)$
Chern-Simons theory on $S^3$. This will be presented in terms of a theorem
which gives the expectation values of Wilson link operators (1.4) in terms of
properties of the plat of a corresponding coloured-oriented braid. This main
theorem and the sketch of its proof has already been announced in ref.17. Here
we are presenting the details of the proof as well as discussing some other
implication of the theorem. For example, a corresponding theorem which
alternatively yields
the link invariants in terms of closure of oriented-coloured braids will also
be present in section 4. Next in section 5, we shall illustrate how the main
theorem can be used to write down the link invariants. This we do by discussing
a
multi-coloured three component link, the Borromean rings.
In section 6, a few concluding remarks will
be made.  Appendix I will contain explicit formulae for the duality matrices in
terms of $ q$-Racah coefficients of $SU(2)_q$ needed in the main text.
Invariants for knots upto eight crossings and multicoloured two component links
upto seven crossings as given in the tables of Rolfsen$^{18}$ will be listed
in Appendis II.

\vspace{1.5cm}

\noindent 2. {\bf Duality of correlators in $SU(2)_k$ Wess-Zumino
conformal field theory}

\vspace{.7cm}

To develop the solution for $SU(2)$ Chern-Simons theory on $S^3$, we need to
make
use of duality properties$^{19}$ of correlators of $SU(2)_k$ Wess-Zumino
conformal
field theory on an $S^2$.  We now list these properties.

Four-point correlators for primary fields with spins $j_1, j_2, j_3$ and
$j_4$ (such that these combine into an $SU(2)$ singlet) can be
 represented in three equivalent ways.  Two such ways are given by figs.1(a)
and (b).  In the first,
each of pairs of spins $j_1, j_2$ and $j_3, j_4$ is  combined into common spin
$j$ representations according to the fusion rules of the $SU(2)_k$ Wess-Zumino
model.  Then these two spin $j$ representations combine to give singlets.  For
sufficiently large values of $k$, allowed values of $j$ are those given by
group
theory:  max${( \vert j_1 - j_2 \vert, \vert j_3 - j_4 \vert ) }$ ${\le j
\le}$min${ (j_1 + j_2,
j_3 + j_4)}$.  In the second equivalent representation for the four-point
correlators spins ${(j_2, j_3)}$ and ${(j_1, j_4)}$ are first combined into
common intermediate spin $\ell$  representation and then two spin $\ell$
representations yield singlets, with max${(\vert j_2 - j_3 \vert,
\vert j_1 -j_4 \vert )}$ ${ \le }$ $ \ell $ ${ \le}$ min${(j_2 + j_3, j_1 +
j_4)}$ for sufficiently large $k$.  These two
sets of linearly independent but equivalent representations will be called
$\phi_j (j_1 j_2 j_3 j_4)$ and $\phi^{\prime}_{\ell} (j_1 j_2 j_3 j_4)$
respectively.  These are related to each other by duality :
$$
\phi_j (j_1 j_2 j_3 j_4)\,=\, \sum_{\ell} \, a_{j\ell} \fourj {j_1} {j_2}
{j_3} {j_4}\,  \phi^{\prime}_{\ell} (j_1 j_2 j_3 j_4)   \eqno(2.1)
$$

\noindent where the duality matrices $ a_{j\ell} \fourj {j_1} {j_2}
{j_3} {j_4} $ are given$^{15,19,6}$ in terms of $q$-Racah coefficients for
$SU(2)_q$.  We have listed these and some of their useful properties
explicitly in Appendix I.  This fact that these two bases are related by
$q$-Racah coefficients is not surprising.  The representation theory of
for integrable representations of SU(2)$_k$ Wess-Zumino field theory is the
same as that of $SU(2)_q$ with the deformation parameter as $q= exp (2\pi
i/(k+2))$.

The duality transformation (2.1) can be successively applied to obtain duality
properties of higher correlators.  In particular, we shall be interested in the
two equivalent sets of correlators for $2m$ primary fields with spin
assignments
${j_1, j_2, \ldots j_{2m},~~ \phi_{(p;r)} (j_1 j_2 \ldots j_{2m})}$ and
$\phi^{\prime}_{(q;s)} (j_1 j_2 \ldots j_{2m})$ as shown in figs.2(a) and (b)
respectively.  Here indices (p) and (r) collectively represent the spins
${(p_0 p_1 \ldots p_{m-1})}$ and ${(r_1 r_2 \ldots r_{m-3})}$ on the internal
lines respectively as shown in fig 2(a).  Similarly, $(q)\, = {(q_0 q_1 \ldots
q_{m-1})}$ and $(s) \,=\, {(s_1 s_2 \ldots s_{m-3})}$ in fig.2(b).  These two
figures represent two equivalent ways of combining spins ${j_1, j_2 \ldots
j_{2m}}$ into singlets and are related by duality.  This fact we now present in
the form of a theorem :

\vspace{.5cm}

{\bf Theorem 1.} The correlators for $2m$ primary fields with spins
 ${j_1, j_2 \ldots}$ $ j_{2m}$, $(m \ge 2)$ in $SU(2)_k$ Wess-Zumino conformal
field theory on an S$^2$ as shown in figs.2(a) and (b) are related to each
other by

$$
\phi_{(p;r)} (j_1 j_2 \ldots j_{2m})\,=\, \sum_{(q;s)}\, a_{(p;r)(q;s)}
 \sixj {j_1} {j_2} {j_3} {j_4} {\vdots} {j_{2m-1}} {j_{2m}} \,
\phi^{\prime}_{(q;s)} (j_1 j_2 \ldots j_{2m})   \eqno(2.2)
$$

\noindent where the duality matrices are given as products of the basic duality
coefficients for the four-point correlators (2.1) as

$$
a_{(p;r)(q;s)} \sixj {j_1} {j_2} {j_3} {j_4} {\vdots} {j_{2m-1}} {j_{2m}} \, =
\sum_{t_1 t_2 \ldots t_{m-2}}\, \prod_{i=1}^{m-2} \\
\left( a_{t_i p_i}\,
\fourj {r_{i-1}} {j_{2i+1}} {j_{2i+2}} {r_i} \, a_{t_i s_{i-1}}\,
\fourj {t_{i-1}} {q_i} {s_i} {j_{2m}} \right)
$$
$$
\quad\quad\quad\quad\quad\quad\quad\quad\quad \times \prod^{m-2}_{\ell=0}\,
a_{r_{\ell} q_{\ell+1}}
\fourj {t_{\ell}} {j_{2\ell+2}} {j_{2\ell+3}} {t_{\ell+1}}   \eqno(2.3)
$$
\noindent Here $r_0 \equiv p_0,~ r_{m-2} \equiv p_{m-1},~ t_0 \equiv j_1,
{}~t_{m-1} \equiv j_{2m},~ s_0 \equiv q_0$ and $s_{m-2} \equiv q_{m-1}$ and
spins
$\vec{j}_1 + \vec{j}_2 + \ldots + \vec{j}_{2m-1} = \vec{j}_{2m}$ and the spins
meeting at trivalent points in fig.2 satisfy the fusion rules of the $SU(2)_k$
conformal field theory.
\vspace{.3cm}

Using the properties of the matrices $a_{j\ell} \left[ \matrix { j_1 & j_2 \cr
j_3 & j_4} \right]$ as given in Appendix I, we can readily see that the duality
matrices (2.3) satisfy the following orthogonality property :
$$
\sum_{(p;r)} \ a_{(p;r) (q;s)} \left[ \matrix{ j_1 & j_2 \cr~~~~~~~ \vdots \cr
j_{2m-1} & j_{2m}} \right] a_{(p;r)(q';s')} \left[ \matrix{ j_1 & j_2 \cr
{}~~~~~~~\vdots \cr j_{2m-1} & j_{2m}} \right] \ = \ \delta_{(q,q')}
\delta_{(s,s')}
\eqno(2.4)
$$

The proof of Theorem 1 is rather straight forward. It can be developed by
applying the duality transformation (2.1) successively on the $2m$-point
correlators. For example, for 6-point correlators, represented by $\phi_{(p_0
p_1 p_2)}$ and $\phi'_{(q_0 q_1 q_2)}$, are related by a sequence of four
duality transformations each involving four spins at a time as shown in fig.3.
Thus
$$
\phi_{(p_0 p_1 p_2)} (j_1 \ldots j_6) \ = \ \sum_{q_0 q_1 q_2} a_{(p_0 p_1 p_2)
(q_0 q_1 q_2)} \left[ \matrix{ j_1 & j_2 \cr j_3 & j_4 \cr j_5 & j_6} \right]
\phi'_{(q_0 q_1 q_2)} (j_1 \ldots j_6) \eqno(2.5)
$$

\noindent where
$$
a_{(p_0p_1p_2)(q_0q_1q_2)} \left[ \matrix{ j_1 & j_2 \cr j_3 & j_4 \cr j_5 &
j_6} \right] \ = \ \sum_{t_1} a_{t_1p_1} \left[ \matrix{ p_0 & j_3 \cr j_4 &
p_2} \right] a_{p_0 q_1} \left[ \matrix{ j_1 & j_2 \cr j_3 & t_1} \right]
$$
$$
\hspace{7cm}\times a_{p_2 q_2} \left[ \matrix{ t_1 & j_4 \cr j_5 & j_6} \right]
a_{t_1 q_0} \left[
\matrix{ j_1 & q_1 \cr q_2 & j_6 } \right] \eqno(2.6)
$$

\noindent Similarly for 8-point correlators $\phi_{(p_0 p_1 p_2 p_3 ; r_1)}
(j_1 .. j_8)$ and $\phi'_{(q_0 q_1 q_2 q_3 ; s_1)} (j_1 .. j_8)$,
which are related by a sequence of seven four-point duality transformations as
shown in Fig.4, we have
$$
\phi_{(p_0 \ldots p_3 ; r_1)} (j_1 \ldots j_8) \ = \ \sum_{(q;s)} a_{(p_0
\ldots p_3 ; r_1) (q_0 \ldots q_3 ; s_1)} \left[ \matrix{ j_1 & j_2 \cr
{}~~~~~~~\vdots
\cr j_7 & j_8} \right] \phi'_{(q_0 .. q_3 ; s_1)} (j_1 \ldots j_8)
$$

\noindent with
$$
a_{(p_0 \ldots p_3 ; r_1) (q_0 \ldots q_3 ; s_1)} \left[ \matrix{ j_1 & j_2
\cr ~~~~~~~\vdots \cr j_7 & j_8} \right] \ = \ \sum_{t_1 t_2}  a_{t_2 p_2}
\left[
\matrix{ r_1 & j_5 \cr j_6 & p_3} \right] a_{t_1 p_1} \left[
\matrix{ p_0 & j_3 \cr j_4 & r_1} \right]
a_{p_0 q_1} \left[
\matrix{ j_1 & j_2 \cr j_3 & t_1} \right]
$$
$$
\hspace {4.7cm} \times a_{r_1 q_2} \left[ \matrix{ t_1 & j_4 \cr j_5 & t_2}
\right] a_{p_3 q_3}
\left[ \matrix{ t_2 & j_6 \cr j_7 & j_8} \right]
$$
$$
\hspace{6.5cm} \times a_{t_2 s_1} \left[
\matrix{ t_1 & q_2 \cr q_3 & j_8} \right] a_{t_1 q_0} \left[
\matrix{ j_1 & q_1 \cr s_1 & j_8} \right] \eqno(2.7)
$$

\noindent Clearly, in this manner Theorem 1 for arbitrary 2m-point correlators
follows.

These duality properties will be made use of in sec.4 to obtain the solution of
$SU(2)$ Chern-Simons theory. But before we do that, we need to discuss the
other ingredient necessary for our purpose. As stated earlier, this has to do
with the theory of coloured-oriented braids.

\vspace{1.5cm}

\noindent {\bf 3. Coloured-oriented braids}

\vspace{.7cm}

An $n$-braid is a collection of non-intersecting strands connecting $n$ points
on a horizontal plane to $n$ points on another horizontal plane directly below
the first set of $n$ points. The strands are not allowed to go back upwards at
any point in their travel. The braid may be projected onto a plane with the two
horizontal planes collapsing to two parallel rigid rods. The over-crossings and
under-crossings of the strands are to be clearly marked. When all the strands
are identical, we have ordinary braids. The theory of such braids is well
developed$^{10,11}$. However, for our purpose here we need to orient the
individual strands and further distinguish them by putting different colours on
them. We shall represent different colours by different $SU(2)$ spins. Examples
of such braids are drawn in Fig.5. These braids, unlike braids made from
identical strands, have a more general structure than a group. These instead
form a groupoid$^{20}$. Now we shall develop some necessary elements of the
theory of groupoid of such coloured-oriented braids.

A general $n$-strand coloured-oriented braid will be specified by giving
$n$ assignments $\hat{j}_i = (j_i, \epsilon_i),~ i=1,2,...n$ representing the
spin
$j_i$ and orientation $\epsilon_i~~ (\epsilon_i = \pm 1$ for the $i$th strand
going into or
away from the rod) on the $n$ points on the upper rod and another set of $n$
spin-orientation assignments $\hat{\ell}_i = (\ell_i, \eta_i)$ on $n$ points on
the lower rod as shown in fig.6. For a spin-orientation assignment $\hat{j}_i =
(j_i, \epsilon_i)$, we define a conjugate assignment as $\hat{j}_i^\ast = (j_i,
-
 \epsilon_i)$. Then the assignments $(\hat{\ell}_i)$ are just a permutation of
$(\hat{j}_i^\ast)$. The shaded box in the middle of the figure represents a
weaving
pattern with various strands going over and under each other. Such a braid will
be represented by the symbol $ \cal {B} \left( \matrix{\hat{j}_1 & \hat{j}_2
\ldots
\hat{j}_n \cr \hat{\ell}_1 & \hat{\ell}_2 \ldots \hat{\ell}_n} \right)$.

\vspace{.5cm}

{\bf Composition :} Unlike usual braids made from identical strands,
the composition for two arbitrary coloured braids is not always defined. Two
such braids ${ \cal{B}} ^{(1)} \left( \matrix{ \hat{j}_1 & \hat{j}_2 \ldots
\hat{j}_n \cr
\hat{j}'_1 & \hat{j}'_2 \ldots \hat{j}'_n} \right)$ and ${ \cal {B}} ^{(2)}
\left(
\matrix{ \hat{\ell}_1 & \hat{\ell}_2 \ldots \hat{\ell}_n \cr \hat{\ell}'_1 &
\hat{\ell}'_2 \ldots \hat{\ell}'_n} \right)$ can be composed only if the
spin-orientations at the merged rods match, that is, the composition ${ \cal
{B}} ^{(1)}
{ \cal {B}} ^{(2)}$ is defined only if $\hat{j}'_i = \hat{\ell}_i^\ast$ and
composition
$ {\cal {B}} ^{(2)}{ \cal {B}} ^{(1)}$ only if $\hat{j}_i^\ast =
\hat{\ell}'_i$.

\vspace{.5cm}

{\bf Generators :} An arbitrary coloured-oriented braid such as one
shown in fig.6 can be generated by applying a set generators on the trivial (no
entanglement) braids $I \left( \matrix { \hat{j}_1 & \hat{j}_2 \ldots \hat{j}_n
\cr \hat{j}_1^\ast & \hat{j}_2^\ast \ldots \hat{j}_n^\ast} \right)$ shown in
fig.7. Unlike the case of usual ordinary braids, here we have more than one
``identity'' braid due to the different values of spin-orientation assignments
$\hat{j}_1, \hat{j}_2, \ldots \hat{j}_n$ placed on the strands. The set of
$n-1$ generators $B_\ell ,~~ \ell=1,2,\ldots n-1$ are represented in fig.7. By
convention we twist the strands by half-units from below keeping the points on
the upper rod fixed. Thus the generator $B_\ell$ introduces from below a
half-twist in the anti-clockwise direction in the $\ell$th and $(\ell+1)$th
strands. Like in the case of usual ordinary braids, the generators of
coloured-oriented braids satisfy two defining relations :
$$ \begin{array}{ccl}
B_i B_{i+1} B_i & = & B_{i+1} B_i B_{i+1} \\
B_i B_j & = & B_j B_i \quad\quad\quad\quad |i-j| \ge 2 \end{array} \eqno(3.1)
$$

\noindent These relations are depicted diagrammatical in figs.8(a) and (b)
respectively. We shall present a whole class fo new representations of these
generators in the next sec.4. These in turn will finally lead to new link
invariants.

\vspace{.5cm}

{\bf Platting of an oriented-coloured braid:} Like in usual case of
braids$^{11}$, we may ~introduce~ the ~concept of~ platting
of a coloured-oriented braid. Consider a coloured-oriented braid with even
number of strands with
spin- orientation assignments as given by $ \cal {B} \left(
\matrix{ \hat{j}_1 & \hat{j}_1^\ast & \hat{j}_2 & \hat{j}_2^\ast & \ldots &
\hat{j}_m & \hat{j}_m^\ast \cr \hat{\ell}_1 & \hat{\ell}_1^\ast & \hat{\ell}_2
&
\hat{\ell}_2^\ast & \ldots & \hat{\ell}_m & \hat{\ell}_m^\ast } \right)$. The
platting then constitutes of pair wise joining of successive strands
$(2i-1,2i)$,$\quad i=1,2,3,\ldots m$ from above and below as shown in fig.9.
Such a
construction obviously can be defined only for  braids made of even number of
strands with above
given specific spin-orientation assignments. There is a theorem due to Birman
which relates oriented links to plats of ordinary even braids$^{11}$. This
theorem can obviously also be stated in terms of coloured-oriented braids of
our present interest. Thus we state
\vspace{.5cm}

{\bf Theorem 2.} A coloured-oriented link can be represented by a
plat constructed from an oriented-coloured braid $ \cal {B} \pmatrix{
{\hat{j}_1} & {\hat{j}_1^\ast} & \ldots & {\hat{j}_m} & {\hat{j}_m^\ast} \cr
{\hat{l}_1} & {\hat{l}_1^\ast} & \ldots & {\hat{l}_m} & {\hat{l}_m}}$.

\vspace{.3cm}

Clearly, platting of these braids does not provide a unique representation of a
given knot or link.

\vspace{.5cm}

{\bf Closure of an oriented-coloured braid}: In addition to platting, we may
also define the closure of a coloured-oriented
braid. For an $m$-strand braid with spin-orientation assignments as in
$ \cal {B} \pmatrix{ {\hat{j}_1} & {\hat{j}_2} & {\ldots} & {\hat{j}_m} \cr
{\hat{j}_1^\ast} & {\hat{j}_2^\ast} & {\ldots} & {\hat{j}_m^\ast}}$, the
closure
of the braid is obtained by joining the top end of each string to the same
position on the bottom of the braid as shown in fig.10. Clearly, closure is
defined only if the spin-orientation assignments are mutually conjugate at the
same positions on the upper and lower rods. Now there is a theorem due to
Alexander$^{16}$ which relates oriented links with closure of ordinary braids.
This theorem can as well be stated for our coloured-oriented braids :

\vspace{.5cm}

{\bf Theorem 3.} A coloured-oriented link can be represented, though
not uniquely, by the closure of an oriented-coloured braid $\cal {B} \eightj
{\hat{j}_1} {\hat{j}_2} {\ldots} {\hat{j}_m} {\hat{j}_1^\ast} {\hat{j}_2^\ast}
{\ldots} {\hat{j}_m^{\ast}}$.
\vspace{.3cm}

In the following we shall see that Theorem 1 with Theorem 2 or 3 provide a
complete solution to $SU(2)$ Chern-Simons gauge theory on an $S^3$.

\newpage
\noindent {\bf 4. Link invariants from $SU(2)$ Chern-Simons theory.}

\vspace{.7cm}

To develope a method of calculating the expectation value of an arbitrary
Wilson link operator (1.4), consider an $S^3$ with two three-balls removed from
it. This is a manifold with two boundaries, each an $S^2$. Let us place $2m
{}~~(m=2,3\ldots)$ unbraided Wilson lines with spins $j_1, j_2,\ldots j_{2m}$
(such
that all these spins make an $SU(2)$ singlet) going from one boundary to the
other as shown in fig.11. Thus we have put an ``identity'' braid $I
\eightj {\hat{j}_1^\ast} {\hat{j}_2^\ast} {\ldots} {\hat{j}_{2m}^\ast}
{\hat{j}_1} {\hat{j}_2} {\ldots} {\hat{j}_{2m}}$ inside the manifold. An
arbitrary coloured-oriented braid can be generated from this identity by
applying the half-twist (braiding) generators $B_1, B_2, \ldots B_{2m-1}$ on
the
lower boundary. As discussed in sec.1, the Chern-Simons functional integral
over this manifold can be represented by a state in the tensor product of
vector spaces, ${\cal{H}}^{(1)} \otimes {\cal{H}}^{(2)}$, associated with the
two
boundaries, $\sum^{(1)}$ and $\sum^{(2)}$. Convenient basis vectors for these
vector spaces can be taken to correspond to the conformal blocks (eqn.(2.2)),
$\phi_{(p;r)} (j_1 j_2 \ldots j_{2m})$ or equivalently $\phi'_{(q;s)} (j_1 j_2
\ldots j_{2m})$ as shown in fig.2 for the $2m$-point correlators of the
corresponding $SU(2)_k$ Wess-Zumino conformal field theory. We
shall represent these bases for each vector space as $ | \phi_{(p;r)}
(\hat{j}_1 \hat{j}_2
\ldots \hat{j}_{2m})>$  and $ | \phi'_{(q;s)} (\hat{j}_1  \hat{j}_2 \ldots
\hat{j}_{2m})>$ respectively.
For dual vector spaces associated with boundaries with opposite orientation, we
have the dual bases $<\phi_{(p;r)} (\hat{j}_1 \ldots \hat{j}_{2m})|$  and $ <
\phi'_{(q;s)} (\hat{j}_1 \ldots \hat{j}_{2m})|$ . The inner product of these
bases vectors for each of the vector space, ${\cal{H}}^{(1)}$ and
${\cal{H}}^{(2)}$
are normalized so that
$$
<\phi_{(p;r)} (\hat{j}_1^\ast \hat{j}_2^\ast \ldots \hat{j}_{2m}^\ast) |
\phi_{(p';r')} (\hat{j}_1 \hat{j}_2 \ldots \hat{j}_{2m}) > = \delta_{(p)(p')}
\delta_{(r)(r')}
$$
$$
<\phi'_{(q;s)} (\hat{j}_1^\ast \hat{j}_2^\ast \ldots \hat{j}_{2m}^\ast) |
\phi'_{(q';s')} (\hat{j}_1 \hat{j}_2 \ldots \hat{j}_{2m}) > = \delta_{(q)(q')}
\delta_{(r)(r')} \eqno(4.1)
$$

The two primed and unprimed bases are related by duality of the
conformal blocks given by Theorem 1 :
$$
| \phi'_{(q;s)} (\hat{j}_1 \hat{j}_2 \ldots \hat{j}_{2m}) = \sum_{(p;r)}
a_{(p;r)(q;s)} \sixj {j_1} {j_2} {j_3} {j_4} {\vdots} {j_{2m-1}} {j_{2m}} \,
| \phi_{(p;r)} (\hat{j}_1 \hat{j}_2 \ldots \hat{j}_{2m})>   \eqno(4.2)
$$

\noindent with duality matrices as in eqn.(2.3).

The Chern-Simons functional integral over the three-manifold of fig.11 may now
be written in terms of one kind the above bases :
$$
\nu_I \eightj {\hat{j}_1^\ast} {\hat{j}_2^\ast} {..} {\hat{j}_{2m}^\ast}
{\hat{j}_1} {\hat{j}_2} {..} {\hat{j}_{2m}} = \sum_{(p;r) (p';r')}
M_{(p;r)(p';r')} | \phi^{(1)}_{(p;r)} (\hat{j}_1^\ast ..
\hat{j}_{2m}^\ast)> | \phi^{(2)}_{(p';r')} (\hat{j}_1 ..
\hat{j}_{2m})>
$$

\noindent Here we have put superscripts (1) and (2) on the bases vectors to
indicate explicitly that they belong to the vector spaces ${\cal{H}}^{(1)}$ and
${\cal{H}}^{(2)}$ respectively. Now notice glueing two copies of this manifold
along two oppositely oriented boundaries, each an $S^2$, yields the same
manifold. Hence
$$
\sum_{(p';r')} M_{(p;r)(p';r')} M_{(p';r') (p'';r'')} = M_{(p;r)(p'';r'')}
$$

\noindent This immediately leads to $M_{(p;r)(p';r')} = \delta_{(p)(p')}
\delta_{(r)(r')}$, so that the functional integral over the three-manifold of
fig.11 can be written as
$$
\nu_I \eightj {\hat{j}_1^\ast} {\hat{j}_2^\ast} {\ldots} {\hat{j}_{2m}^\ast}
{\hat{j}_1} {\hat{j}_2} {\ldots} {\hat{j}_{2m}} = \sum_{(p;r)}
| \phi_{(p;r)}^{(1)} (\hat{j}_1^\ast \ldots \hat{j}_{2m}^\ast)> |
\phi^{(2)}_{(p;r)} (\hat{j}_1 \ldots \hat{j}_{2m})> \eqno(4.3a)
$$

\noindent Equivalently we could write this functional integral in terms of the
primed basis using (4.2) and orthogonality property of duality matrices (2.4)
as
$$
\nu_I \left( \matrix { {\hat{j}_1^\ast} & {\ldots} & {\hat{j}_{2m}^\ast} \cr
{\hat{j}_1} & {\ldots} & {\hat{j}_{2m}}} \right) = \sum_{(q;s)}
| \phi'^{(1)}_{(q;s)} (\hat{j}_1^\ast \ldots \hat{j}_{2m}^\ast)> |
\phi'^{(2)}_{(q;s)} (\hat{j}_1 \ldots \hat{j}_{2m})> \eqno(4.3b)
$$

The conformal blocks $\phi_{(p;r)} (j_1 \ldots j_{2m})$ as shown in fig.2(a) of
the conformal field theory and the corresponding basis vectors  $|
\phi_{(p;r)} (\hat{j}_1 \hat{j}_2 \ldots \hat{j}_{2m})>$ are eigen functions of
the~ odd indexed ~braiding ~generators ~~$B_{2\ell+1},~~ \ell =0,1 \ldots
(m-1)$ of
fig.7.~~ On the~ other~ hand ~the ~~conformal~~ blocks $ \phi'_{(q;s)} ({j}_1
{j}_2
\ldots {j}_{2m})$ (fig.2b) and the associated basis vector  $ |\phi'_{(q;s)}
(\hat{j}_1 \hat{j}_2  \ldots \hat{j}_{2m})>$ are eigen functions of the even
indexed braid generators, $B_{2\ell},~~ \ell=1,2,\ldots (m-1)$ :
$$
B_{2\ell+1} |\phi_{(p;r)} (\ldots \hat{j}_{2\ell+1} \hat{j}_{2\ell+2} \ldots)>
= \lambda_{p\ell} (\hat{j}_{2\ell+1} , \hat{j}_{2\ell+2}) | \phi_{(p;r)}
(\ldots \hat{j}_{2\ell+2} \hat{j}_{2\ell+1} \ldots)>
$$
$$
B_{2\ell} |\phi'_{(q;r)} (\hat{j}_1 ..  \hat{j}_{2\ell} \hat{j}_{2\ell+1}
 .. \hat{j}_{2m})> = \lambda_{q\ell} (\hat{j}_{2\ell} , \hat{j}_{2\ell+1}) |
\phi_{(q;s)} (\hat{j}_1 ..  \hat{j}_{2\ell+1} \hat{j}_{2\ell} ..
\hat{j}_{2m})>  \eqno(4.4)
$$

\noindent The eigenvalues of the half-twist matrices depend on the relative
orientation of the twisted strands :
$$ \begin{array}{lclcll}
\lambda_t (\hat{j}, \hat{j}') & = & \lambda^{(+)}_t (j,j') & \equiv &
(-)^{j+j'-t} q^{(c_j + c_{j'})/2 + c_{min(j,j')} - c_t/2} & \quad {\rm if} \
\epsilon
\epsilon ' = +1 \\
& = & (\lambda^{(-)}_t (j,j'))^{-1} & \equiv & (-)^{|j-j'|-t} q^{|c_j -
c_{j'}|/2 - c_t/2} & \quad {\rm if} \ \epsilon \epsilon ' = -1 \end{array}
\eqno(4.5)
$$

\noindent where $c_j = j(j+1)$ is the quadratic Casimir for spin $j$
representation. When $\epsilon \epsilon' = +1$ above, the two-strands have the
same
orientation and the braid generator introduces a right-handed half-twist as
shown in fig.12a. On the other hand for $\epsilon \epsilon' = -1$, the two
strands are
anti-parallel and the braid generator introduces a left-handed half-twist as
shown in fig.12b. Thus $\lambda^{(+)}_t (j,j')$ and $\lambda_t^{(-)} (j,j')$
above are the eigenvalues of the half-twist matrix which introduce right-handed
half-twists in parallely and anti-parallely oriented strands respectively.
These eigen-values are obtained from the monodromy properties of the conformal
blocks of fig.2 of the corresponding conformal theory$^{19} $ and further
compensated
for the change of framing introduced due to the twisting of the
strands$^{3,6}$. There is some ambiguity with regard to the $q$-independent
phases in these
expressions for the eigenvalues. However, this ambiguity along with that in
the phase of the duality matrix $a_{j\ell}$ of eqn.(2.1) are relatively fixed
by consistency requirements as will be discussed in the Appendix I below.

Eqns.(4.4), (4.5) and (4.2) define representations of braids. This we express
in the form of a theorem :

\vspace{.3cm}

{\bf Theorem 4} . A class of representations for generators of the groupoid of
coloured-oriented braids of fig.7 are given (in the basis $| \phi_{(p;r)}>$) by
$$
\left[ B_{2\ell+1} \left( \matrix{ \hat{j}_1^\ast & .. &
\hat{j}_{2\ell+1}^\ast & \hat{j}_{2\ell+2}^\ast & .. & \hat{j}_{2m}^\ast
\cr .. & .. & \hat{j}_{2\ell+2} & \hat{j}_{2\ell+1} & .. & ..}
\right) \right]_{(p;r)(p';r')}   =  \lambda_{p_\ell}
(\hat{j}_{2\ell+1}, \hat{j}_{2\ell+2}) \delta_{(p)(p')} \delta_{(r)(r')}
$$
$$
 \hspace {9.8cm} \ell = 0,1,\ldots (m-1)$$
\newpage
and
$$
\left[ B_{2\ell} \left( \matrix{ \hat{j}_1^\ast & .. &
\hat{j}_{2\ell}^\ast & \hat{j}_{2\ell+1}^\ast & .. & \hat{j}_{2m}^\ast \cr
\hat{j}_1 & .. & \hat{j}_{2\ell+1} & \hat{j}_{2\ell} & .. &
\hat{j}_{2m}} \right) \right]_{(p;r)(p';r')} \hspace{5cm}
$$
$$
\hspace{1cm}  =  \sum_{(q;s)} \lambda_{q_\ell}
(\hat{j}_{2\ell}, \hat{j}_{2\ell+1}) ~a_{(p';r')(q;s)} \left[ \matrix{ j_1 &
j_2
\cr \vdots & \vdots \cr j_{2\ell-1} & j_{2\ell+1} \cr j_{2\ell} & j_{2\ell+2}
\cr \vdots & \vdots \cr j_{2m-1} & j_{2m}} \right]
 a_{(p;r)(q;s)} \left[ \matrix{ j_1 & j_2
\cr \vdots & \vdots \cr j_{2\ell-1} & j_{2\ell} \cr j_{2\ell+1} & j_{2\ell+2}
\cr \vdots & \vdots \cr j_{2m-1} & j_{2m}} \right]
$$
$$
\hspace{7cm} \ell = 1, 2, \ldots (m-1)  \eqno(4.6)
$$
\vspace{.3cm}

\noindent Using the identities given in Appendix I, these can readily be
verified to satisfy the defining relations (3.3) of the braid generators.

	Now let us place an arbitrary weaving pattern instead of an identity braid
inside the three-manifold with two boundaries (each an $ S^2 $) discussed above
with specific spin-orientation assignments as shown in fig.13. The
spin-orientation assignment $(\hat{\ell}_1, \hat{\ell}_2 \ldots \hat{\ell}_m)$
on
the lower boundary are just a permutations of $(\hat{j}_1^\ast, \hat{j}_2^\ast
\ldots \hat{j}_m^\ast)$. The braid inside indicated as a shaded box can be
represented in terms of a word ${\cal B}$ in the braid generators $B_i$ above.
The
Chern-Simons functional integral over this three-manifold can thus be obtained
by ${\cal B}$ (written in terms generators $B_i$) acting on the state (4.3)
from below
:
$$
\nu_{\cal B} \left( \matrix{ {\hat{j}_1} & {\hat{j}_1^\ast} & \ldots &
{\hat{j}_m} &
{\hat{j}_m^\ast} \cr {\hat{l}_1} & {\hat{l}_1^\ast} & \ldots & {\hat{l}_m} &
{\hat{l}_m}} \right) = \sum_{(p;r)} | \phi^{(1)}_{(p;r)}>~~ {\cal B} |
{}~\phi^{(2)}_{(p;r)} >  \eqno(4.7)
$$

We wish to plat this braid. This can be done by glueing one copy each of the
three-ball shown in fig.14(a) from below and above with spin-orientation
assignments matching at the punctures. The functional integral over this
three-ball (fig.14(a)) can again be thought of to be a vector in the Hilbert
space
associated with the boundary. Thus we write the functional integral (normalized
by multiplying by $Z^{-1/2}$ where $Z$ is the partition function on $S^3$) in
terms of a basis of this Hilbert space as
$$
\nu(\hat{j}_1 \hat{j}_1^\ast \ldots \hat{j}_m \hat{j}_m^\ast) = \sum_{(p;r)}
N_{(p;r)} ~~| \phi_{(p;r)} (\hat{j}_1 \hat{j}_1^\ast \ldots \hat{j}_m
\hat{j}_m^\ast) >
$$

\noindent where the coefficients $N_{(p;r)}$ are to be fixed. Notice applying
an arbitrary combination of odd indexed braid generators $B_{2\ell+1}$ on
fig.14(a) does not change this manifold;~ the ~half-twists~ so~ introduced ~can
{}~simply be~ undone. ~That shows ~that ~the ~vector~~ $\nu(\hat{j}_1
\hat{j}_1^\ast
\ldots \hat{j}_m \hat{j}_m^\ast)$~~ is ~proportional to ~~$|\phi_{(0;0)}
(\hat{j}_1
\hat{j}_1^\ast \ldots \hat{j}_m \hat{j}_m^\ast)>$ which is the eigen-function
of the generators $B_{2\ell+1}$ with eigenvalue one. Thus only non-zero
coefficient is $N_{(0;0)}$. Further if we glue two copies of the three-ball of
fig.14(a) onto each other along their oppositely oriented boundaries, we obtain
an $S^3$ containing $m$ unlinked unknots carrying spins $j_1,j_2,\ldots j_m$
respectively. The invariant for this link is given simply by the product of
invariants for individual unknots. Now for cabled knots such as two unknots,
the invaraints satisfy the fusion rules of the associated conformal field
theory. Thus for unknots $V_{j_1}[U] V_{j_2} [U] = \sum_{j} V_j[U]$ where the
spins $(j_1 , j_2 , j)$ are related by the fusion rules of the conformal field
theory. For spin $1/2$ representation using skein relations we can obtain
 ~ $V_{1/2} [U] =[2]$, where square brackets define $q$-numbers as
$[x]=(q^{x/2}
- e^{-x/2})/(q^{1/2} - q^{-1/2})$. Using this along with $V_0 [U]=1$, the
invariant for unknot $U$ can be seen to be given by the $q$-dimension of the
representation living on the knot $V_j[U] = [2j+1]$. This discussion leads to
$N_{(0;0)} = \prod  [2j_i +1]^{1/2}$ above. Thus
$$
\nu(\hat{j}_1 \hat{j}_1^\ast \ldots \hat{j}_m \hat{j}_m^\ast) = (\prod^m_{i=1}
[2j_i+1]^{1/2}) | \phi_{(0;0)} (\hat{j}_1 \hat{j}_1^\ast \ldots \hat{j}_m
\hat{j}_m^\ast)> \eqno(4.8)
$$

Now we are ready to plat the braid in the manifold of fig.13 by glueing to it
manifolds of the type shown in fig.14(a) from below and above. This, invoking
Theorem 2, leads us to our main theorem :

\vspace{.3cm}

{\bf Theorem 5.} The expectation value (1.4) of a Wilson operator for an
arbitrary link $L$ with a plat representation in terms of a coloured-oriented
braid ${\cal B} \left( \matrix{ {\hat{j}_1} & {\hat{j}_1^\ast} & \ldots &
{\hat{j}_m} & {\hat{j}_m^\ast} \cr {\hat{l}_1} & {\hat{l}_1^\ast} & \ldots &
{\hat{l}_m} & {\hat{l}_m^\ast}} \right)$ generated by a word written in terms
of
the braid generators $B_i,~~ i=1, 2,\ldots (2m-1)$, is given by
$$
V[L] = (\prod^m_{i=1} [2j_i+1]) \hspace{10cm}
$$
$$
\hspace{1.8cm}\times  < \phi_{(0;0)}({\hat \ell}_1^\ast {\hat \ell}_1 ..
\hat{\ell}_m^\ast \hat{\ell}_m) | {\cal B} \left( \matrix{ {\hat{j}_1} &
{\hat{j}_1^\ast} & .. & {\hat{j}_m} & {\hat{j}_m^\ast} \cr {\hat{\ell}_1} &
{\hat{\ell}_1^\ast} & .. & {\hat{\ell}_m} & {\hat{\ell}_m^\ast}} \right) |
\phi_{(0;0)} (\hat{j}_1^\ast \hat{j}_1 .. \hat{j}_m^\ast \hat{j}_m)> \eqno(4.9)
$$
\vspace{.3cm}

This main theorem along with Theorem 4 allows us to calculate the link
invariant for any arbitrary link. Before illustrating this with an explicit
example in the next section, we shall extend our discussion developed above to
write down the Chern-Simons functional integral over a three-ball with Wilson
lines as shown in fig.14(b). One way to obtain this functional integral is by
applying a weaving pattern generated by $B_{2m-1} B_{2m} \ldots B_3 B_2$ on the
functional integral (4.8) for the three-ball shown in fig.14(a). Alternatively,
since this functional integral is unchanged by applying even-indexed braid
generators $B_{2\ell}$, it is proportional to $ | \phi'_{(0;0)} (\hat{j}_1
\hat{j}_2 \hat{j}_2^\ast \ldots \hat{j}_m \hat{j}_m^\ast \hat{j}_1^\ast)>$
which is the eigen function of these generators with eigenvalue one.  This
functional integral over the ball of fig.14(b) (normalized by multiplying by
$Z^{-1/2}$) is given by the vector
$$
\nu'(\hat{j}_1 \hat{j}_2 \hat{j}_2^\ast \ldots \hat{j}_m \hat{j}_m^\ast
\hat{j}_1^\ast) = (-)^{2j_1}(\prod^m_{i=1} (-)^{2min(j_1,j_i)} [2j_i+1]^{1/2})
{}~| \phi'_{(0;0)} (\hat{j}_1
\hat{j}_2 \hat{j}_2^\ast \ldots \hat{j}_m \hat{j}_m^\ast \hat{j}_1^\ast) >
\eqno(4.10)
$$

Similarly the Chern-Simons functional integral for the three-ball of fig.14(c)
can be constructed by applying the braid $g_{2m} \equiv (B_{m+1} B_m)$$
(B_{m+2}
B_{m+1}$ $ B_m$ $ B_{m-1})$
$(B_{m+3} B_{m+2} \ldots B_{m-2}) \ldots (B_{2m-1} \ldots B_3 B_2)$ on the
vector (4.8) representing functional integral
over the manifold of fig.14(a). We write
$$
|\tilde{\phi} (\hat{j}_1 \hat{j}_2 \ldots \hat{j}_m \hat{j}_m^\ast \ldots
\hat{j}_2^\ast \hat{j}_1^\ast)> = g_{2m}~ | \phi_{(0;0)} (\hat{j}_1
\hat{j}_1^\ast \hat{j}_2 \hat{j}_2^\ast \ldots \hat{j}_m \hat{j}_m^\ast)>
\eqno(4.11)
$$
\noindent Then the Chern-Simons normalized functional integral over this
three-ball is
$$
\tilde{\nu} (\hat{j}_1 \hat{j}_2 .. \hat{j}_m \hat{j}_m^\ast ..
\hat{j}_2^\ast \hat{j}_1^\ast) = (\prod^m_{i=1}  [2j_i+1]^{1/2})~ |
\tilde{\phi} (\hat{j}_1 \hat{j}_2 ..  \hat{j}_m \hat{j}_m^\ast
.. \hat{j}_2^\ast \hat{j}_1^\ast) > \eqno(4.12)
$$

This ~functional ~integral allows us ~to~~ obtain a ~result ~equivalent ~to
Theorem 5
for the links as ~represented ~by closure ~of braids. ~To do so, ~~for ~an
{}~braid
${\cal B}_m \eightj  {\hat{j}_1} {\hat{j}_2} {..} {\hat{j}_m} {\hat{j}_1^\ast}
{\hat{j}_2^\ast} {..} {\hat{j}_m^\ast}$ with~ $m$ strands ,~construct ~~another
{}~~~braid ~~by adding ~~$m$ ~untangled ~~strands ~to ~obtain a ~$2m$-strand~~
braid ${\hat {\cal B}}_{m}
\pmatrix{ \hat{j}_1 & \hat{j}_2 & \ldots & \hat{j}_m & \hat{j}_m^\ast & \ldots
& \hat{j}_2^\ast & \hat{j}_1^\ast \cr  \hat{j}_1^\ast & \hat{j}_2^\ast & \ldots
& \hat{j}_m^\ast & \hat{j}_m & \ldots & \hat{j}_2 & \hat{j}_1}$ as shown in
fig.15 with the spin-orientation assignments as indicated. Then the closure of
the original $m$-strand braid ${\cal B}_m$ in $S^3$ is obtained by glueing two
copies, one
each from above and below, of the three-ball of fig.14(c) onto the manifold of
fig.15 with proper matching of spin-orientations on the punctures on the
boundaries. Thus, we may state the result  for links represented as closure of
braids as :

\vspace{.3cm}

{\bf Theorem 6.} \ For a link represented by the closure of an $m$-strand
coloured-oriented braid ${\cal B}_m \eightj  {\hat{j}_1} {\hat{j}_2} {\ldots}
{\hat{j}_m} {\hat{j}_1^\ast} {\hat{j}_2^\ast} {\ldots} {\hat{j}_m^\ast}$, the
link invaraint is given in terms of the extended $2m$-braid ${ \hat{\cal
B}}_{m}
\pmatrix{ \hat{j}_1 & \hat{j}_2 & \ldots & \hat{j}_m & \hat{j}_m^\ast & \ldots
& \hat{j}_2^\ast & \hat{j}_1^\ast \cr  \hat{j}_1^\ast & \hat{j}_2^\ast & \ldots
& \hat{j}_m^\ast & \hat{j}_m & \ldots & \hat{j}_2 & \hat{j}_1}$ constructed by
adding $m$ untangled strands as in fig.15, by
$$
V[L] = (\prod^m_{i=1} [2j_i+1]) <\tilde{\phi} (\hat{j}_1 \hat{j}_2 ..
\hat{j}_m \hat{j}_m^\ast .. \hat{j}_2^\ast \hat{j}_1^\ast) | {\hat {\cal
B}}_{m} \pmatrix{ \hat{j}_1 & \hat{j}_2 & .. & \hat{j}_m & \hat{j}_m^\ast & ..
& \hat{j}_2^\ast & \hat{j}_1^\ast \cr  \hat{j}_1^\ast & \hat{j}_2^\ast & ..
& \hat{j}_m^\ast & \hat{j}_m & .. & \hat{j}_2 & \hat{j}_1}
$$
$$
\hspace{6cm}  |\tilde{\phi} \pmatrix{ \hat{j}_1^\ast  \hat{j}_2^\ast
..  \hat{j}_m^\ast  \hat{j}_m  .. \hat{j}_2 \hat{j}_1 }>
\eqno(4.13)
$$
\noindent Here the $2m$-strand braid is written as a word in terms of the braid
generators $ B_1,  B_2,   \ldots  B_{m-1}$  introducing weaving pattern in the
first $m$ strands only and the vector
$|\tilde{\phi}>$ is given by eqn.(4.11) above.

\vspace{.3cm}

Theorem 5 or equivalently Theorem 6 provides a complete and
explicit solution of $SU(2)$ Chern-Simons gauge theory on $S^3$.

\newpage
\noindent {\bf 5. Applications of the main theorem}

\vspace{.5cm}

To illustrate the use of the main Theorem 5, let us calculate the invariant for
Borromean rings. This link is made from three knots. We shall place spin $j_1,
j_2$ and $j_3$ on these knots. Fig.16 shows this links with orientation and
spin assignments as indicated. A plat representation for this link has also
been
drawn. The link is given as a plat of a six strand braid $ B_2  B_4^{-1} B_3
B_1
 B_4^{-1} B_3 B_2^{-1} B_4^{-1}$. To apply Theorem
5, first we evaluate $ B_2^{-1} B_4^{-1} | \phi_{(0)}>$.
This we do by first converting the basis vector $| \phi_{(0)}>$ to $|
\phi'_{(\ell)}>$ through duality matrix and since $ B_2^{-1}
B_4^{-1}$ introduces right-handed half twists in anti-parallel strands:
$$
 B^{-1}_2  B^{-1}_4 | \phi_{(0)} (\hat{j}_2 \hat{j}_2^\ast \hat{j}_1
\hat{j}_1^\ast \hat{j}_3 \hat{j}_3^\ast)>  \hspace{8.5cm}
$$
$$
\hspace {1cm} = \sum_{(\ell_i)} \lambda^{(-)}_{\ell_1} (j_1 j_2)
\lambda^{(-)}_{\ell_2} (j_1 j_3) a_{(0)(\ell)} \left[ \matrix{j_2 & j_2 \cr j_1
& j_1 \cr j_3 & j_3} \right]~ | \phi'_{(\ell)} (\hat{j}_2 \hat{j}_1
\hat{j}_2^\ast \hat{j}_3
\hat{j}_1^\ast \hat{j}_3^\ast ) >
$$
\noindent Next we apply $B_3$ ( which introduces a left-handed half-twist in
the
anti-parallel strands) on this vector. For this we change the basis back to
$ |\phi_{(m)} >$ through duality transformation. Repeating such steps, we
finally have using Theorem 5, the invariant for the Borromean rings of fig.16
as
$$
V_{j_1 j_2 j_3} \ = \   [2 j_1 +1][2 j_2 + 1][2 j_3 + 1] \hspace{9cm}
$$
$$\hspace{1.8cm}\times < \phi_{(0)} ( \hat{j}_1^\ast \hat{j}_1 \hat{j}_2^\ast
\hat{j}_2
\hat{j}_3^\ast \hat{j}_3 ) | ~ B_2 B^{-1}_4 B_3 B_1 B^{-1}_4 B_3 B^{-1}_2
B^{-1}_4 ~|\phi_{(0)} ( \hat{j}_2 \hat{j}_2^\ast \hat{j}_1 \hat{j}_1^\ast
\hat{j}_3 \hat{j}_3^\ast ) >
$$
$$\hspace{1.2cm} \ =\  [2 j_1 +1][2j_2 +1][2j_3 +1] \sum \left(
\lambda^{(-)}_{q_1} (j_1 j_2) \right)^{-1}  \lambda^{(-)}_{q_2} (j_2 j_3 )
\lambda^{(+)}_{p_0} (j_1 j_2 )
$$
$$\hspace{3cm} \times~\left( \lambda^{(-)}_{p_1} (j_1 j_3) \right)^{-1}
 \left( \lambda^{(+)}_{n_2} (j_1 j_2)
\right)^{-1} \left(\lambda^{(-)}_{m_1} (j_2 j_3)\right)^{-1}
\lambda^{(-)}_{\ell_1} (j_1 j_2) \lambda^{(-)}_{\ell_2} (j_1 j_3)
$$
$$
\hspace{4cm} \times a_{(0)(q)} \left[ \matrix{ j_1 & j_1 \cr j_2 & j_2 \cr j_3
& j_3}
\right] a_{(p)(q)} \left[ \matrix{ j_1 & j_2 \cr j_1 & j_3 \cr j_2 & j_3}
\right] a_{(p)(n)} \left[ \matrix{ j_2 & j_1 \cr j_3 & j_1 \cr j_2 & j_3}
\right]
$$
$$
\hspace{4cm} \times a_{(m)(n)} \left[ \matrix{ j_2 & j_1 \cr j_3 & j_2 \cr j_1
& j_3}
\right] a_{(m)(\ell)} \left[ \matrix{ j_2 & j_1 \cr j_2 & j_3 \cr j_1 & j_3}
\right] a_{(0)(\ell)} \left[ \matrix{ j_2 & j_2 \cr j_1 & j_1 \cr j_3 & j_3}
\right] \eqno(5.1)
$$

Similarly the knot invariants for example, for all the knots and links list in
the tables given in Rolfsen's book$^{18}$ may be calculated. We shall present
the result of such calculations for knots upto eight crossings and
two-component links upto seven crossings in Appendix II. Some of these
invaraints were calculated earlier in refs.6.

\vspace{1cm}

\noindent {\bf 6. Concluding remarks}

\vspace{.7cm}

We have here presented an explicit method for obtaining the functional average
of an arbitrary Wilson link operator (1.4) in an $SU(2)$ Chern-Simons theory.
Either of the main Theorems 5 or 6 provides this complete solution. To develop
this method, we have made use of theory of coloured-oriented braids. In
addition, following Witten$^{3}$, we have used the equivalence of the Hilbert
space of Chern-Simons functional integrals over a three-manifold with boundary
with the vector space of the conformal blocks for the correlators of the
associated Wess-Zumino conformal field theory on that boundary based on the
same group and same level. This has helped us to find a whole class of new
representations of generators of coloured-oriented braids. These in turn have
finally led to the explicit solution of the Chern-Simons gauge theory. Of the
new link invariants so obtained, Jones polynomial is the simplest. It
corresponds to spin $1/2$ representation living on all the components of the
link. The new invaraints appear to be more powerful than Jones polynomial as
these do distinguish knots which are known to have the same Jones polynomials.

Tables of the new invaraints for knots and links of low crossing numbers have
been presented in Appendix II. We could read off the invaraints for any
other links as well by the rules defined by Theorem 5 or 6. In particular,
invaraints for toral knots can be obtained in this way.

Theorems 5 and 6 can also be used for an efficient calculation of the
invariants
on a computer.

The method developed has an obvious generalizations to other compact
semi-simple gauge groups. It can also be extended to study Chern-Simons gauge
theory on three-manifolds other than $S^3$.

\vspace {2cm}

\noindent {\bf Appendix I}

\vspace{.7cm}

Here we list the duality matrix $a_{j\ell}\left[ \matrix{ j_1 & j_2 \cr j_3 &
j_4} \right]$ relating the two bases of four-point correlators of $SU(2)_k$
Wess-Zumino field theory as shown in fig.1. We shall also give some of their
useful properties. These duality matrices are given in terms of $q$-Racah
coefficients as $^{19,6}$ :
$$
{a_{j\ell}} \left[ \matrix{ j_1 & j_2 \cr j_3 & j_4} \right] =
(-)^{(j_1 + j_2 + j_3 + j_4)} \sqrt{[2j+1][2\ell +1]} \left( \matrix{
j_1 & j_2 & j \cr j_3 & j_4 & \ell} \right) \eqno(A.1)
$$
Here the triplets$ (j j_1 j_2)$, $(j j_3 j_4)$, $(\ell j_1 j_4)$ and $(\ell
j_2j_3)$ satisfy the fusion rules of the conformal theory:
$$
max(|j_1-j_2|, |j_3-j_4|) \ge j \ge  min(j_1+j_2, j_3+j_4)
$$
$$
max(|j_2-j_3|, |j_1-j_4|) \ge \ell \ge  min(j_2+j_3, j_1+j_4)
$$
$$
j_1+j_2+j \le k, ~~ j_3+j_4+j \le k, ~~j_2+j_3+ \ell \le k, ~~j_1+j_4+ \ell \le
k
$$
$$
j_1+j_2+j,~~j_3+j_4+j,~~j_2+j_3+\ell~~and~~ j_1+j_4+ \ell  ~~\in  {\bf Z}
\eqno(A.2)
$$
\noindent The phase in (A.1) is so chosen that it is real; $ (j_1+j_2
+j_3+j_4)$ is always an integer.

The $SU(2)_q$ Racah-Wigner coefficients$^{15}$  are :
$$
\left( \matrix{ j_1 & j_2 & j_{12} \cr j_3 & j_4 & j_{23}} \right) = \Delta
(j_1,j_2, j_{12}) \Delta (j_3,j_4,j_{12}) \Delta (j_1,j_4,j_{23}) \Delta
(j_2,j_3,j_{23}) \hspace{3cm}
$$
$$ \begin{array}{ll}
& \hspace{2.7cm} \times \sum_{m \ge 0} (-)^m [m+1]! ~~ {\bf \{ } [m-j_1 - j_2 -
j_{12}]! \\
&\hspace{3.8cm} \times [m - j_3 - j_4 - j_{12}]! [m - j_1 - j_4 - j_{23}]!  \\
& \hspace{3.8cm} \times [m - j_2 - j_3 - j_{23}]! [j_1 + j_2 + j_3 + j_4 - m]!
\\
&\hspace{3.8cm}  \times  [j_1+j_3+j_{12}+j_{23}-m]! [j_2+j_4+j_{12}+j_{23}-m]!
 { \bf \} } ^{-1}
\end{array} \eqno(A.3)
$$
where
$$
\Delta(a,b,c) = \sqrt{ \frac{[-a+b+c]! [a-b+c]! [a+b-c]!} {[a+b+c+1]!}}
$$

\noindent Here the square brackets represent the $q$-numbers
$$
[x] = \frac{q^{x/2}-q^{-x/2}}{q^{1/2}-q^{-1/2}}
$$

\noindent and $[n]! = [n][n-1] \ldots [3][2][1]$. The $SU(2)$ spins are related
as $\vec{j}_1 + \vec{j}_2 + \vec{j}_3 = \vec{j}_4, \vec{j}_1 + \vec{j}_2
= \vec{j}_{12},  \vec{j}_2 + \vec{j}_3 = \vec{j}_{23}$.

The $q$-Racah coefficients above satisfy the following properties$^{15}$ :
Interchange
of any two columns of $\left( \matrix{ j_1 & j_2 & j \cr j_3 & j_4 & \ell}
\right)$ leaves it unchanged. Further
$$
\left( \matrix{ j_1 & j_2 & j \cr j_3 & j_4 & \ell} \right) = \left( \matrix{
j_1 & j_4 & \ell \cr j_3 & j_2 & j} \right) = \left( \matrix{ j_3 & j_2 & \ell
\cr j_1 & j_4 & j} \right) = \left( \matrix{ j_3 & j_4 & j \cr j_1 & j_2 &
\ell}\right) \eqno(A.4)
$$
$$
\left( \matrix{ j_1 & j_2 & 0 \cr j_3 & j_4 & \ell} \right) =
\frac{(-)^{\ell+j_2+j_3}
\delta_{j_1j_2} \delta_{j_3j_4}}{\sqrt{[2j_2+1] [2j_3+1]}} \eqno(A.5)
$$
$$
\sum_j [2j+1][2\ell+1] \left( \matrix{ j_1 & j_2 & j \cr j_3 & j_4 & \ell}
\right)  \left( \matrix{ j_1 & j_2 & j \cr j_3 & j_4 & \ell'} \right)  =
\delta_{\ell\ell'}
\eqno(A.6)
$$
$$
\sum_x (-)^{j+\ell+x} [2x+1] q^{-C_{x/2}} \left( \matrix{ j_1 & j_2 & x \cr j_3
& j_4 & j} \right)  \left( \matrix{ j_1 & j_2 & x \cr j_4 & j_3 & \ell} \right)
\hspace{6cm}
$$
$$
 \hspace{3cm} = \left( \matrix{ j_3 & j_2 & j \cr j_4 & j_1 & \ell} \right)
q^{C_{j/2} + C_{\ell/2}}
q^{-C_{j_1/2} - C_{j_2/2} -C_{j_3/2} - C_{j_4/2}} \eqno(A.7)
$$
$$
\sum_{\ell_1} (-)^{\ell_1 +\ell_2 +\ell_3 + r_1 +r_2} [2\ell_1 +1] \left(
\matrix{r_1 & j_3 & r_2 \cr j_4 & j_5 & \ell_1 } \right)
\quad \left( \matrix{ j_1 & j_2 & r_1 \cr \ell_1 & j_5 & \ell_2 } \right) \quad
\left( \matrix{ \ell_2 & j_2 & \ell_1 \cr j_3 & j_4 & \ell_3 } \right)
\hspace{6cm}
$$
$$ \hspace{3cm} \ = \ (-)^{j_1+j_2+j_3+j_4+j_5}
\left( \matrix{ j_1 & \ell_3 & r_2 \cr j_4 & j_5 & \ell_2 } \right) \quad
\left( \matrix{j_1 & j_2 & r_1 \cr j_3 & r_2 & \ell_3 } \right)    \eqno(A.8)
$$
\noindent where $C_j = j(j+1)$.

Using these we see that the duality matrices satisfy the orthogonality and
symmetry properties as :
$$
\sum_{j} ~~ a_{j\ell} \left[
\matrix{ j_1 & j_2 \cr j_3 & j_4 } \right] a_{j\ell'} \left[\matrix{ j_1 & j_2
\cr j_3 & j_4 } \right] = \delta_{\ell\ell'} \eqno(A.9)
$$
\vspace{.3cm}
$$
a_{j\ell} \left[ \matrix{ j_1 & j_2 \cr j_3 & j_4} \right] = a_{\ell j}
\left[ \matrix{ j_1 & j_4 \cr j_3 & j_2} \right] = a_{\ell j} \left[
\matrix{ j_3 & j_2 \cr j_1 & j_4} \right] = a_{j\ell} \left[ \matrix{ j_3 & j_4
\cr j_1 & j_2} \right] \eqno(A.10)
$$

\noindent Further
$$
a_{j\ell} \left[\matrix{j_1 & j_2 \cr j_3 & j_4} \right] = a_{j\ell} \left[
\matrix{ j_2 & j_1 \cr j_4 & j_3}\right] = (-)^{j_1+j_3 -j -\ell} ~a_{j_1j_2}
\left[\matrix{j & j_2 \cr \ell & j_4} \right]  \eqno(A.11)
$$
$$
a_{0\ell} \left[ \matrix{ j_1 & j_2 \cr j_3 & j_4} \right] = (-)^{j_1 +j_3
-\ell}
\sqrt{ \frac{[2\ell+1]} {[2j_2+1][2j_3+1]}} ~\delta_{j_1 j_2} \delta_{j_3 j_4}
 ;\quad a_{j \ell} \left[ \matrix{ 0 & j_2 \cr j_3 & j_4} \right] = \delta_{j_2
j} \delta_{j_4 \ell} \eqno(A.12)
$$
and
$$
(-)^{2min(j_1,j_2)} (\lambda_m (\hat{j}_1 \hat{j}_2))^{\pm1} a_{m0}
\left[\matrix{j_2 & j_1 \cr j_1 & j_2} \right] = \sum_{\ell} a_{0\ell} \left[
\matrix{j_1 & j_1 \cr j_2 & j_2}
\right] ( \lambda_{\ell} ( \hat{j}_1 \hat{j}_2^{\ast}))^{\mp1} a_{m\ell}
\left[ \matrix{ j_1 & j_2 \cr j_1 & j_2} \right] \eqno(A.13)
$$
$$
{\sum_{m \ell}}~a_{ms} \left[ \matrix{j_3 & j_2 \cr j_1 & j_4} \right]
\lambda_m (\hat{j}_2 \hat{j}_3 ) a_{m\ell} \left[\matrix{j_2 & j_3 \cr j_1 &
j_4 } \right] \lambda_{\ell} (\hat{j}_1 \hat{j}_3)a_{p\ell} \left[\matrix{j_2 &
j_1 \cr j_3 & j_4}\right] \
lambda_p (\hat{j}_1 \hat{j}_2) \hspace{2cm}
$$
$$
\hspace{.7cm} = \sum_{m\ell} \lambda_s(\hat{j}_1 \hat{j}_2 )a_{ms}
\left[\matrix{ j_3 & j_1 \cr j_2 & j_4} \right] \lambda_m (\hat{j}_1 \hat{j}_3)
 a_{m\ell} \left[ \matrix{ j_1 & j_3 \cr j_2 & j_4} \right] \lambda_{\ell}
( \hat{j}_2 \hat{j}_3 ) a_{p \ell} \left[ \matrix{ j_1 & j_2 \cr j_3 & j_4}
\right] \eqno(A.14)
$$
$$
\sum_{\ell_1}~a_{r_2\ell_1}\left[\matrix{r_1 & j_3 \cr j_4 & j_5
}\right]a_{r_1\ell_2}\left[\matrix{j_1 & j_2 \cr \ell_1 &
j_5}\right]a_{\ell_1\ell_2}\left[\matrix{\ell_2 & j_2 \cr j_3 & j_4}\right]
\hspace{4cm}
$$
$$ \hspace{3cm} = a_{r_2\ell_2}\left[\matrix{j_1 & \ell_3 \cr j_4 & j_5}\right]
a_{r_1\ell_3}\left[\matrix{j_1 & j_2 \cr j_3 & r_2}\right] \eqno(A.15)
$$
\noindent Eqn. (A.14)~ reflects~~ the ~~generating ~~relation of ~the
{}~~braiding ~generators $B_i B_{i+1} B_i$ ~= $~B_{i+1} B_i B_{i+1} $ . Both
(A.13) and (A.14) follow
immediately from the applications of the identity (A.7).

The $ q$-independent phases in the eigenvalues $ \lambda^{(\pm)}_{\ell} (j_1
j_2)$ of the braiding matrices given in eqn. (4.5) and also that in the duality
matrix (A.1) are somewhat ambiguous. The choice we make here is  differs from
that in refs.6.
We have chosen these phases in such a way that $ \lambda^{(-)}_0 (j, j) =
1$,~$\lambda^{(\pm)}_{\ell} (0,j) = \delta_{\ell j}$ and $ a_{j \ell} \left[
\matrix{0 & j_2
\cr j_3 & j_4} \right] $ =  $ a_{j \ell} \left[ \matrix{ j_2 & 0 \cr j_4 & j_3
}\right] $ = $ a_{j \ell} \left[ \matrix{j_3 & j_4 \cr 0 & j_2} \right] $ =
$a_{j \ell} \left[ \matrix{j_4 & j_3 \cr j_2 & 0} \right] $ = $ \delta_{j j_2}
 \delta_{\ell j_4} $. The braiding relation (A.14) is not sensitive to this
ambiguity of phases. However, these phases are relatively fixed by requiring
some consistency conditions. One such condition is obtained by gluing a copy
each of the manifolds shown in figs.14(a) and (b) for $m=2$ (and with
$j_1=j_2=j$)
along their oppositely oriented boundaries. This yields an unknot $U$ carrying
spin $j$ in an $S^3$. Thus we have, using eqns.(4.8,10) for $m=2$ and
$j_1=j_2$,
 the consistency condition:
$$
[2j+1]^2 (-)^{2j} <\phi_{0}(\hat{j} \hat{j}^{\ast} \hat{j} \hat{j}^{\ast}) |
\phi_{0}
^{\prime} (\hat{j}^{\ast} \hat{j} \hat{j}^{\ast} \hat{j})>  =  [2j +1]^2
(-)^{2j}~a_{00}
 \left[\matrix{j & j \cr j & j}\right]
=  V_j [U]    \eqno(A.16)
$$
Another consistency condition is eqn. (A.13). This reflects the equality of
Chern-Simons functional integrals over two three-balls as shown in fig.17(a). A
weaker condition on $\lambda^{(\pm)}_{\ell} (j_1,j_2)$ is
$$
\sum_{\ell} [2 \ell+1] \left( \lambda^{(+)}_{\ell} (j_1 j_2) \right)^{\pm2 \ }
= \  \sum_{\ell} [2\ell+1] \left( \lambda^{(-)}_{\ell} (j_1 j_2) \right)^{\pm2}
  \eqno(A.17)
$$
\noindent This condition is obtained by gluing two copies each of the diagrams
of fig. 17(a) to represent the same Hopf link in two different ways. Yet
another consistency condition is
$$
 [2j+1]^{2} \sum_{\ell} \ a_{0\ell} \left[ \matrix{j & j \cr j & j }
\right]~a_{0\ell} \left[ \matrix{j & j \cr j & j } \right] \  \left(
\lambda^{(+)}_{\ell} (jj) \right)^{\pm1} \   = \
{}~[2j + 1]    \eqno(A.18)
$$
\noindent This equation represents the fact that each of the knots in fig.17(b)
is an unknot $ U $ .

Next, eqn.(A.15) follows directly from eqn. (A.8). This equation reflects that
five duality transformations on conformal blocks of the conformal field
theory as shown in fig.18 bring us back to the same block. However a phase may
be picked up in the process. With the phase of duality matrices as fixed above,
there is no such phase picked up by this cycle of five duality transformations.

The duality matrices for some low values of $j$ can easily computed explicitly
. For example,
$$
a_{j\ell} \left[ \matrix{1/2 & 1/2 \cr 1/2 & 1/2} \right] \ = \ {{1}\over{[2]}}
\left(\matrix{-1 & \sqrt{[3]} \cr \sqrt{[3]} & 1} \right) \hspace{2.5cm}
\eqno(A.19)
$$
and
$$
a_{j\ell}\left[\matrix{1 & 1 \cr 1 & 1}\right] \ = \ {1
\over{[3]}}~\left(\matrix{1 & {-\sqrt{[3]}} & {\sqrt{[5]}} \cr
{-\sqrt{[3]}} & {{[3]([5]-1)}\over{[4][2]}} & {{[2]\sqrt{[5][3]}}\over{[4]}}
\cr
\sqrt{[5]} & {{[2]\sqrt{[5][3]}}\over{[4]}} & {[2]\over{[4]}} } \right)
\eqno(A.20)
$$

The general duality matrix $a_{(p;r) (q:s)}$ for $2m$-point correlators of
Wess-Zumino conformal theory as depicted in figs.2 are given by eqn.(2.3) of
Theorem 1. Using above four-point duality matrices, these can be shown to
satisfy the orthogonality and symmetry properties expressed in eqns.(2.4)
. Further the some special values of these $2m$-point duality matrices are
$$
{a_{(0;0) (q;s)}} \left[ \matrix{ jj \cr jj \cr \vdots \cr jj} \right] \ = \
\prod^{m-2}_{\ell=0} \left( (-)^{2j-q_{\ell+1}} \ {{\sqrt{ [2q_{\ell+1} +1]}}
\over{[2j+1]}} \right) \\
\prod^{m-2}_{i=1} a_{j s_{i-1}} \left[ \matrix{ j & q_i \cr s_i & j} \right]
$$
$$
{a_{(p;r) (0;0)}} \left[ \matrix{ jj \cr jj \cr \vdots \cr jj} \right] \ = \
\prod^{m-2}_{\ell=0} \ \left( (-)^{2j- r_{\ell}} {{ \sqrt{[2r_{\ell} +1]}}
\over{[2j+1]}} \right) \\
\prod^{m-2}_{i=1} a_{j p_{i}} \left[ \matrix{ r_{i-1} & j \cr j & r_i} \right]
  \eqno(A.21)
$$
\noindent where $s_0 \equiv q_0, s_{m-2} \equiv q_{m-1}, r_0 = p_0, r_{m-2} =
p_{m-1}$.
\noindent Further a useful identity is :
$$
{{ {\bf \sum } }_{(q_i; i\ne \ell), ( s_j; j \ne \ell, \ell-1)}} a_{(p;r)(q;s)}
\left[ \matrix{ j_1 & j_2 \cr \vdots & \vdots \cr
j_{2\ell-3} & j_{2\ell-2} \cr j_{2\ell-1} & j_{2\ell} \cr j_{2\ell+1} &
j_{2\ell+2} \cr \vdots & \vdots \cr j_{2m-1} & j_{2m}} \right]
a_{(p';r')(q;s)} \left[ \matrix{ j_1 & j_2 \cr \vdots & \vdots \cr
j_{2\ell-3} & j_{2\ell-2} \cr j_{2\ell-1} & j_{2\ell+1} \cr j_{2\ell} &
j_{2\ell+2} \cr \vdots & \vdots \cr j_{2m-1} & j_{2m}} \right] \hspace{7cm}
$$
$$
\hspace{.8cm} = \left( \prod_{i\ne \ell-1,\ell} \delta_{p_i p'_i} \right)
\left( \prod_{j\ne \ell-1,\ell} \delta_{r_j r'_j} \right)
a_{s_{\ell-1} p'_{\ell-1}} \fourj {r_{\ell-2}} {j_{2\ell-1}} {j_{2\ell+1}}
{r'_{\ell-1}}
$$
$$\hspace{1.6cm}\times  a_{s_{\ell-1} p_{\ell-1}} \fourj {r_{\ell-2}}
{j_{2\ell-1}}
{j_{2\ell}} {r_{\ell-1}}
a_{s_{\ell} p'_{\ell}} \fourj {r'_{\ell-1}} {j_{2\ell}} {j_{2\ell+2}}
{r_{\ell}}
$$
$$ \hspace{3cm}\times a_{s_{\ell} p_{\ell-1}} \fourj {r_{\ell-1}} {j_{2\ell+1}}
{j_{2\ell+2}} {r_{\ell}}
a_{r'_{\ell-1} q_{\ell}} \fourj {s_{\ell-1}} {j_{2\ell+1}} {j_{2\ell}}
{s_{\ell}} a_{r_{\ell-1}} q_{\ell} \fourj {s_{\ell-1}} {j_{2\ell}}
{j_{2\ell+1}} {s_{\ell}}
$$
$$
\hspace{7cm} \ell \ = \ 0,1,2,\ldots (m-1)\,. \eqno(A.18)
$$
\noindent Here $r_{-1} \equiv 0$.

\vspace{1.5cm}

\noindent {\bf Appendix II}

\vspace{.5cm}

It may be worthwhile to present a tabulation of the new invariants for knots
and links. This we present now for knots and links with low crossing numbers as
listed by tables of Rolfsen$^{19}$. The naming of knots and links will be given
as in this book which reads clearly the crossing number (as the minimal number
of
double points in the link diagram). We shall not present the link diagrams as
shown in these tables but instead give their plat representations so that
Theorem 5 can readily be used to write down the invariants.

\vspace{.5cm}

{\bf IIA. Knots:} In this subsection invariants for knots upto crossing number
eight will be given. All knots will carry spin $j$ representations and we shall
shorten the notation for eigen values of the braid matrix introducing
right-handed half-twists in parallely and antiparallely oriented strands and
also
the duality matrices as :
$$
\lambda_{\ell}^{(\pm)} ~ \equiv ~ \lambda^{(\pm)}_\ell (j,j)  , \quad\quad\quad
a_{m\ell}
{}~\equiv ~a_{m\ell} \left[ \matrix{ jj \cr jj}  \right],
$$
$$
a_{(m)(\ell)} \ \equiv \ a_{(m_0 m_1 m_2) (\ell_0 \ell_1 \ell_2)} \left[
\matrix{ jj \cr jj \cr jj} \right]
$$

The plat representation of knots studied here are given in fig.19 and their
knot invariants $V_j $ using Theorem 5 are as follows :

\vspace{.5cm}

\noindent $0_1 : \qquad [2j+1]$

\vspace{.5cm}

\noindent $3_1 : \qquad \sum [2\ell+1] (\lambda^{(+)}_\ell)^{-3}$

\vspace{.5cm}

\noindent $ 4_1 : \qquad \sum \sqrt{[2m+1] [2\ell+1]}(-)^{m+\ell-2j}
(\lambda^{(-)}_m)^{-2}
(\lambda_\ell^{(-)})^2 a_{m\ell}$

\vspace{.5cm}

\noindent $ 5_1 : \qquad \sum [2\ell+1] (\lambda^{(+)}_\ell)^{-5}$

\vspace{.5cm}

\noindent $ 5_2 : \qquad \sum \sqrt{[2m+1] [2\ell+1]}
(-)^{m+\ell-2j}(\lambda^{(-)}_m)^{-3}
(\lambda_\ell^{(+)})^{-2} a_{m\ell}$

\vspace{.5cm}

\noindent $6_1 : \qquad \sum \sqrt{[2m+1] [2\ell+1]}
(-)^{m+\ell-2j}(\lambda^{(-)}_m)^{-4}
(\lambda_\ell^{(-)})^{2} a_{m\ell}$

\vspace{.5cm}

\noindent $ 6_2 : \qquad \sum \sqrt{[2p+1] [2m+1]} (-)^{p+m}
(\lambda^{(+)}_p)^{-3}
(\lambda_n^{(-)})^{-1}  (\lambda^{(-)}_m)^2 a_{pn} a_{mn}$

\vspace{.5cm}

\noindent $ 6_3 : \qquad \sum \sqrt{[2p+1] [2\ell+1]}(-)^{p+\ell-2j}
(\lambda^{(+)}_p)^{2}
(\lambda_n^{(-)}) (\lambda^{(-)}_m)^{-1} (\lambda^{(+)}_\ell)^{-2} a_{pn}
a_{mn}
a_{m\ell}$

\vspace{.5cm}

\noindent $7_1 : \qquad \sum [2\ell +1] (\lambda^{(+)}_\ell)^{-7}$

\vspace{.5cm}

\noindent $ 7_2 : \qquad \sum \sqrt{[2m+1] [2\ell+1]}
(-)^{m+\ell-2j}(\lambda^{(-)}_m)^{-5}
(\lambda_\ell^{(+)})^{-2} a_{m\ell}$

\vspace{.5cm}

\noindent $7_3 : \qquad \sum \sqrt{[2m+1] [2\ell+1]} (-)^{m+\ell-2j}
(\lambda^{(+)}_m)^{4}
(\lambda_\ell^{(-)})^{3} a_{m\ell}$

\vspace{.5cm}

\noindent $7_4 : \qquad \sum \sqrt{[2p+1] [2m+1]} (-)^{p+m}
(\lambda^{(-)}_p)^{3}
(\lambda_n^{(+)}) (\lambda^{(-)}_m)^{3} a_{pn} a_{mn}$

\vspace{.5cm}

\noindent $7_5 : \qquad \sum \sqrt{[2p+1] [2m+1]} (-)^{p+m}
(\lambda^{(+)}_p)^{-3}
(\lambda_n^{(-)})^{-2}  (\lambda^{(+)}_m)^{-2} a_{pn} a_{mn}$

\vspace{.5cm}

\noindent $7_6 : \qquad \sum \sqrt{[2p+1] [2\ell+1]} (-)^{p+\ell-2j}
(\lambda^{(+)}_p)^{-2}
(\lambda_n^{(-)})^{-1} (\lambda^{(-)}_m)^{2} (\lambda^{(-)}_\ell)^{-2} a_{pn}
a_{mn} a_{m\ell}$

\vspace{.5cm}

\noindent $7_7 : \qquad \sum \sqrt{[2r+1] [2m+1]} (-)^{r+m}
(\lambda^{(-)}_r)^{2}
(\lambda_q^{(-)})^{-1} (\lambda^{(+)}_p)^{-1} (\lambda^{(-)}_n)^{-1}
(\lambda^{(-)}_m)^2  a_{rq} a_{pq} a_{pn} a_{mn}$

\vspace{.5cm}

\noindent $8_1 : \qquad \sum \sqrt{[2m+1] [2\ell+1]} (-)^{m+\ell-2j}
(\lambda^{(-)}_m)^{-6}
(\lambda_\ell^{(-)})^{2} a_{m\ell}$

\vspace{.5cm}

\noindent $8_2 : \qquad \sum \sqrt{[2p+1] [2m+1]} (-)^{p+m}
(\lambda^{(+)}_p)^{-5}
(\lambda_n^{(-)})^{-1}  (\lambda^{(-)}_m)^{2} a_{pn} a_{mn}$

\vspace{.5cm}

\noindent $8_3 : \qquad \sum \sqrt{[2m+1] [2\ell+1]} (-)^{m +\ell-2j}
(\lambda^{(-)}_m)^{-4}
(\lambda_\ell^{(-)})^{4} a_{m\ell}$

\vspace{.5cm}

\noindent $8_4 : \qquad [2j+1]^3 \sum (\lambda^{(-)}_{n_1})^3
(\lambda^{(+)}_{n_2})^{-3} (\lambda^{(+)}_{m_0}) (\lambda^{(+)}_{m_1})
(\lambda^{(-)}_{\ell_1} \lambda^{(+)}_{\ell_2})^{-1}$ \\
\indent\indent\indent\indent\indent\indent $\times  a_{(0)(n)}
a_{(m)(n)} a_{(m)(\ell)} a_{(0)(\ell)}$

\vspace{.5cm}

\noindent $8_5 : \qquad [2j+1]^3 \sum \lambda^{(-)}_{n_1}
(\lambda^{(+)}_{n_2})^{-1}
(\lambda^{(+)}_{m_0}) (\lambda^{(+)}_{m_1})^3
\lambda^{(-)}_{\ell_1} (\lambda^{(+)}_{\ell_2})^{-1}$ \\
\indent\indent\indent\indent\indent\indent $\times  a_{(0)(n)}
a_{(m)(n)} a_{(m)(\ell)} a_{(0)(\ell)}$

\vspace{.5cm}

\noindent $8_6 : \qquad \sum \sqrt{[2p+1] [2m+1]} (-)^{p+m}
(\lambda^{(+)}_p)^{-3}
(\lambda_n^{(-)})^{-3} (\lambda^{(-)}_m)^{2} a_{pn} a_{mn} a_{m\ell}$

\vspace{.5cm}

\noindent $8_7 : \qquad \sum \sqrt{[2p+1] [2\ell+1]} (-)^{p+\ell-2j}
(\lambda^{(+)}_p)^{4}
(\lambda_n^{(-)}) (\lambda^{(-)}_m)^{-1} (\lambda^{(+)}_\ell)^{-2} a_{pn}
a_{mn} a_{m\ell}$

\vspace{.5cm}

\noindent $8_8 : \qquad [2j+1]^3 \sum  (\lambda^{(-)}_{n_1})^{-1}
(\lambda^{(+)}_{n_2})^{2} (\lambda^{(+)}_{m_1})^{-1} (\lambda^{(-)}_{m_2})^2
(\lambda^{(-)}_{\ell_1})^{-1} \lambda^{(+)}_{\ell_2}$ \\
\indent\indent\indent\indent\indent\indent $ \times a_{(0)(n)} a_{(m)(n)}
a_{(m)(\ell)} a_{(0)(\ell)}$

\vspace{.5cm}

\noindent $8_9 : \qquad \sum \sqrt{[2p+1] [2\ell+1]} (-)^{p+\ell-2j}
(\lambda^{(+)}_p)^{3}
\lambda_n^{(-)} (\lambda^{(-)}_m)^{-1} (\lambda^{(+)}_\ell)^{-3} a_{pn}
a_{mn} a_{m\ell}$

\vspace{.5cm}

\noindent $8_{10} : \qquad [2j+1]^3 \sum  (\lambda^{(+)}_{n_1})^{3}
(\lambda^{(-)}_{n_2})^{-1} (\lambda^{(+)}_{m_1})^{-1}
(\lambda^{(+)}_{\ell_1})^2 (\lambda^{(-)}_{\ell_2})^{-1}$ \\
\indent\indent\indent\indent\indent\indent $ \times a_{(0)(n)} a_{(m)(n)}
a_{(m)(\ell)} a_{(0)(\ell)}$

\vspace{.5cm}

\noindent $8_{11} : \qquad [2j+1]^3 \sum  (\lambda^{(-)}_{n_1})^{-1}
(\lambda^{(+)}_{n_2}) (\lambda^{(+)}_{m_0})^{-1} (\lambda^{(-)}_{m_1})^{-1}
(\lambda^{(-)}_{\ell_1})^{-3} \lambda^{(+)}_{\ell_2}$ \\
\indent\indent\indent\indent\indent\indent $\times a_{(0)(n)} a_{(m)(n)}
a_{(m)(\ell)} a_{(0)(\ell)}$

\vspace{.5cm}

\noindent $8_{12} : \qquad \sum \sqrt{[2p+1] [2\ell+1]} (-)^{p+\ell-2j}
(\lambda^{(-)}_p)^{-2}
(\lambda_n^{(-)})^2 (\lambda^{(-)}_m)^{-2} (\lambda^{(-)}_\ell)^{2} a_{pn}
a_{mn} a_{m\ell}$

\vspace{.5cm}

\noindent $8_{13} : \qquad \sum \sqrt{[2r+1] [2\ell+1]} (-)^{r+\ell}
(\lambda^{(-)}_r)^{3}
(\lambda_q^{(+)}) (\lambda_p^{(-)}) (\lambda^{(-)}_n)^{-1}
(\lambda^{(+)}_m)^{-2} a_{rq} a_{pq} a_{pn} a_{mn} $

\vspace{.5cm}

\noindent $8_{14} : \qquad \sum \sqrt{[2q+1] [2\ell+1]} (-)^{q+\ell}
(\lambda^{(-)}_q)^{2}
(\lambda_p^{(-)})^{-1} (\lambda_n^{(+)})^{-1} (\lambda^{(-)}_m)^{-2}
(\lambda^{(+)}_\ell)^{-2} a_{pq} a_{pn} a_{mn} a_{m\ell}$

\vspace{.5cm}

\noindent $8_{15} : \qquad [2j+1]^3 \sum (\lambda^{(-)}_{q_2})^{-1}
(\lambda^{(+)}_{p_2})^{-1} (\lambda^{(-)}_{n_1} \lambda^{(-)}_{n_2})^{-1}
(\lambda^{(+)}_{m_1})^{-1} (\lambda^{(-)}_{\ell_1})^{-1}
(\lambda^{(-)}_{\ell_2})^{-2}$ \\
\indent \indent \indent\indent \indent\indent $\times a_{(0)(q)}
a_{(p)(q)}a_{(p)(n)} a_{(m)(n)}
a_{(m)(\ell)} a_{(0)(\ell)}$

\vspace{.5cm}

\noindent $8_{16} : \qquad [2j+1]^3 \sum (\lambda^{(-)}_{q_1}
\lambda^{(-)}_{q_2}) ^{-1}
\lambda^{(-)}_{p_1} (\lambda^{(+)}_{p_2})^{-1} \lambda^{(+)}_{n_2}
\lambda^{(-)}_{m_1} {(\lambda^{(-)}_{\ell_1} \lambda^{(-)}_{\ell_2})}^{-1}$ \\
\indent\indent\indent\indent \indent \indent $\times a_{(0)(q)} a_{(p)(q)}
a_{(p)(n)} a_{(m)(n)}
a_{(m)(\ell)} a_{(0)(\ell)}$

\vspace{.5cm}

\noindent $8_{17} : \qquad [2j+1]^3 \sum (\lambda^{(-)}_{s_2})^{-1}
(\lambda^{(+)}_{r_1})^{-1} \lambda^{(-)}_{q_1} (\lambda^{(+)}_{p_0})^{-1}
\lambda^{(-)}_{p_1} \lambda^{(+)}_{n_1} \lambda^{(+)}_{n_2}
\lambda^{(-)}_{m_1} (\lambda^{(-)}_{\ell_1} \lambda^{(-)}_{\ell_2})^{-1}$ \\
\indent \indent \indent\indent\indent\indent $\times a_{(0)(s)} a_{(r)(s)}
a_{(r)(q)} a_{(p)(q)}
a_{(p)(n)} a_{(m)(n)} a_{(m)(\ell)} a_{(0)(\ell)}$

\vspace{.5cm}

\noindent $ 8_{18} : \qquad [2j+1]^3 \sum \lambda^{(-)}_{q_1}
\lambda^{(-)}_{q_2}
(\lambda^{(-)}_{p_1})^{-1} (\lambda^{(+)}_{n_1} \lambda^{(+)}_{n_2})^{-1}
(\lambda^{(-)}_{m_1})^{-1} \lambda^{(-)}_{\ell_1} \lambda^{(-)}_{\ell_2}$ \\
\indent \indent \indent\indent\indent\indent $\times a_{(0)(q)} a_{(p)(q)}
a_{(p)(n)} a_{(m)(n)}
a_{(m)(\ell)} a_{(0)(\ell)}$

\vspace{.5cm}

\noindent $8_{19} : \qquad [2j+1]^3 \sum \lambda^{(-)}_{q_1}
\lambda^{(-)}_{q_2}
\lambda^{(+)}_{p_0} \lambda^{(-)}_{p_1} \lambda^{(+)}_{n_1}
\lambda^{(+)}_{n_2} (\lambda^{(+)}_{m_0})^{-1} \lambda^{(-)}_{m_1}
\lambda^{(-)}_{\ell_1} \lambda^{(-)}_{\ell_2}$ \\
\indent\indent\indent\indent \indent \indent $\times a_{(0)(q)} a_{(p)(q)}
a_{(p)(n)} a_{(m)(n)}
a_{(m)(\ell)} a_{(0)(\ell)}$

\vspace{.5cm}

\noindent $8_{20} : \qquad [2j+1]^3 \sum (\lambda^{(-)}_{q_1})^{-1}
\lambda^{(-)}_{q_2}
(\lambda^{(+)}_{p_0} \lambda^{(-)}_{p_1})^{-1} \lambda^{(+)}_{n_1}
(\lambda^{(+)}_{n_2})^{-1} (\lambda^{(+)}_{m_0})^{-1} \lambda^{(-)}_{m_1}
\lambda^{(-)}_{\ell_1} (\lambda^{(-)}_{\ell_2})^{-1}$ \\
\indent \indent \indent\indent\indent\indent $\times a_{(0)(q)} a_{(p)(q)}
a_{(p)(n)} a_{(m)(n)}
a_{(m)(\ell)} a_{(0)(\ell)}$

\vspace{.5cm}

\noindent $8_{21} : \qquad [2j+1]^3 \sum (\lambda^{(-)}_{q_1}
\lambda^{(-)}_{q_2}) ^{-1}
(\lambda^{(+)}_{p_0})^{-1} \lambda^{(-)}_{p_1} ( \lambda^{(+)}_{n_1}
\lambda^{(+)}_{n_2})^{-1} (\lambda^{(+)}_{m_0} \lambda^{(-)}_{m_1})^{-1}
\lambda^{(-)}_{\ell_1} \lambda^{(-)}_{\ell_2}$ \\
\indent \indent \indent\indent\indent\indent $\times a_{(0)(q)} a_{(p)(q)}
a_{(p)(n)} a_{(m)(n)}
a_{(m)(\ell)} a_{(0)(\ell)}$

\vspace{.5cm}

In the expressions above all the indices are summed over positive integers
from $0$ to min$(2j, k-2j)$ . In these calculations we have made
use of identities (A.12) and (A.13). The results for knots upto crossing number
seven as presented in the first of ref.6 are the same. Further these invaraints
for $j=1/2$ and $j=1$ respectively agree with Jones-one variable
polynomials$^{2}$ and those obtained by Wadati {\it et al} from the three-state
exactly solvable model$^{14}$. To do this comparison we need to multiply these
polynomials of refs.$2$ and $14$ by $[2j+1]$,~~ $j=1/2$ and $1$ respectively,
to account for differences of normalization before comparing with our results
above.

Notice for $q = 1$ (which corresponds to $k \rightarrow \infty$ ), the
invariant for any knot is
simply the ordinary dimension of the representation living on it,
$V_j (q=1) \ = \ 2j+1$.

Change of orientation does not affect the knots and invariants do not depend on
the orientation. Thus orientation for knots may not be specified. However,
mirror reflected knots are not isotopically equivalent in general. For any
chiral knot in the above list, the invariant for the obverse is obtained by
conjugation which amounts to replacing various braid matrix eigen values
$\lambda^{(\pm)}_\ell$ by their inverses in the expression.

\vspace{.5cm}

{\bf II B. Links :} \ Now we shall list the invariants for two component links
with crossing number upto seven as listed in Rolfsens book$^{18}$. Unlike in
the case
of knots above, here we need to specify the orientations on the two components.
There are four possible ways of puting arrows on these knots. Simultaneous
reversing of
orientations on all the component knots does not change the
invariant. Hence there are only two independent ways of specifying the
orientations on the knots of a two component link. We have made a specific
choice of relative orientations of the component knots as indicated in fig.20
where we have given plat representations of these links. We have also placed
spin $j_1$ and $j_2$ representations on the components as indicated. Then from
Theorem 5, the invariants $V_{j_1 j_2}$ for these links are :

\vspace{.5cm}

\noindent $0_1 : \qquad [2j_1+1] [2j_2+1]$

\vspace{.5cm}

\noindent $2_1 : \qquad \sum [2\ell+1] (\lambda^{(+)}_\ell (j_1 j_2))^{-2}$

\vspace{.5cm}

\noindent $4_1 : \qquad \sum [2\ell+1] (\lambda^{(+)}_\ell (j_1 j_2))^{-4}$

\vspace{.5cm}

\noindent $5_1 : \qquad [2j_1+1]^2 [2j_2+1] \sum (\lambda^{(-)}_{p_1} (j_1
j_2))^{-1}
\lambda^{(+)}_{p_2} (j_1 j_2) (\lambda^{(+)}_{m_1} (j_1 j_1))^{-1}$ \\
 \indent\indent\indent\indent\indent\indent\indent\indent$ \times
(\lambda^{(-)}_{\ell_1} (j_1 j_2))^{-1} \lambda^{(+)}_{\ell_2} (j_1 j_2)
a_{(0)(p)} \left[ \matrix{ j_1 j_1 \cr j_2 j_2 \cr j_1 j_1} \right]$ \\
\indent\indent\indent\indent\indent\indent\indent\indent $\times  a_{(m)(p)}
\left[ \matrix{ j_1 j_2 \cr j_1 j_1
\cr j_2 j_1} \right] a_{(m)(\ell)} \left[ \matrix{ j_1 j_2 \cr j_1 j_1
\cr j_2 j_1} \right] a_{(0)(\ell)} \left[ \matrix{ j_1 j_1 \cr j_2 j_2
\cr j_1 j_1} \right]$

\vspace{.5cm}

\noindent $6_1 : \qquad \sum [2\ell+1] (\lambda^{(+)}_{\ell} (j_1 j_2))^{-6}$

\vspace{.5cm}

\noindent $6_2 : \qquad \sum \sqrt{[2\ell+1] [2m+1]} (-)^{m+\ell-2max(j_1,j_2)}
(\lambda^{(-)}_{m} (j_1 j_2)
\lambda^{(+)}_{\ell} (j_1 j_2))^{-3} a_{(m)(\ell)} \left[ \matrix{ j_1 j_2 \cr
j_1 j_2 } \right]$

\vspace{.5cm}

\noindent $6_3 : \qquad  \sum \sqrt{[2p+1] [2m+1]} (-)^{p+m-2(j_1+j_2)}
(\lambda^{(-)}_{p} (j_1 j_2))^{2}
(\lambda^{(-)}_{n} (j_1 j_2))^{-2}(\lambda^{(-)}_{m} (j_1 j_2))^{2} $ \\
\indent\indent\indent\indent\indent\indent$\times
a_{pn} \left[ \matrix{ j_1 j_2 \cr j_2 j_1 } \right] a_{mn} \left[ \matrix{ j_1
j_2 \cr j_2 j_1 } \right]$

\vspace{.5cm}

\noindent $7_1 : \qquad  \sum \sqrt{[2p+1] [2m+1]} (-)^{p+m-2(j_1+j_2)}
(\lambda^{(-)}_{p} (j_1 j_2))^{4}
(\lambda^{(-)}_{n} (j_2 j_2))^{-1}(\lambda^{(+)}_{m} (j_1 j_2))^{-2}$ \\
\indent\indent\indent\indent\indent\indent$\times
a_{pn} \left[ \matrix{ j_1 j_2 \cr j_2 j_1 } \right] a_{mn}  \left[ \matrix{
j_1 j_2 \cr
j_2 j_1 } \right]$

\vspace{.5cm}

\noindent $7_2 : \qquad  \sum \sqrt{[2p+1] [2\ell+1]}
(-)^{p+\ell-2max(j_1,j_2)} (\lambda^{(-)}_{p} (j_1 j_2))^{-3}
(\lambda^{(+)}_{n} (j_1 j_2))^{-1} $ \\
\indent\indent\indent\indent\indent\indent $ (\lambda^{(-)}_{m} (j_2 j_2))^{-1}
(\lambda^{(-)}_{\ell} (j_1 j_2))^{2}$  \\
\indent \indent \indent\indent\indent\indent $ \times a_{pn} \left[ \matrix{
j_1 j_2 \cr j_1 j_2 }  \right]
a_{mn} \left[ \matrix{ j_1 j_1 \cr j_2 j_2 } \right] a_{m\ell} \left[ \matrix{
j_1 j_1 \cr j_2 j_2 } \right]$

\vspace{.5cm}

\noindent $7_3 : \qquad  \sum \sqrt{[2p+1] [2m+1]} (-)^{p+m-2(j_1+j_2)}
(\lambda^{(-)}_{p} (j_1
j_2))^{2} (\lambda^{(-)}_{n} (j_2 j_2))^{-3} (\lambda^{(+)}_{m} (j_1
j_2))^{-2}$ \\
\indent\indent\indent\indent\indent\indent$\times
a_{pn} \left[ \matrix{ j_1 j_2 \cr j_2 j_1 }  \right] a_{mn} \left[ \matrix{
j_1
j_2 \cr j_2 j_1 } \right]$

\vspace{.5cm}

\noindent $7_4 : \qquad [2j_1+1] [2j_2+1]^2 \sum \lambda^{(-)}_{n_1} (j_1 j_2)
\lambda^{(-)}_{n_2} (j_2 j_2) (\lambda^{(-)}_{m_1} (j_1 j_2))^{-2}$ \\
\indent\indent\indent\indent\indent\indent$\times
\lambda^{(+)}_{m_2} (j_2 j_2) \lambda^{(-)}_{\ell_1} (j_1 j_2)
\lambda^{(-)}_{\ell_2} (j_2 j_2)$ \\
\indent\indent\indent \indent\indent\indent\indent \indent $\times a_{(0)(m)}
\left[ \matrix{ j_1 j_1 \cr j_2 j_2
\cr j_2 j_2} \right] a_{(m)(n)} \left[ \matrix{ j_1 j_2 \cr j_1 j_2
\cr j_2 j_2} \right] a_{(m)(\ell)} \left[ \matrix{ j_1 j_2 \cr j_1 j_2
\cr j_2 j_2} \right] a_{(0)(\ell)} \left[ \matrix{ j_1 j_1 \cr j_2 j_2
\cr j_2 j_2} \right]$

\vspace{.5cm}

\noindent $7_5 : \qquad [2j_1+1] [2j_2+1]^2 \sum (\lambda^{(-)}_{q_1} (j_1 j_2)
\lambda^{(-)}_{q_2} (j_1 j_2)) (\lambda^{(-)}_{p_1} (j_2 j_2))^{-1}
\lambda^{(+)}_{p_2} (j_1 j_2) (\lambda^{(+)}_{n_2} (j_2 j_2))^{-1}$ \\
\indent\indent\indent \indent\indent\indent\indent \indent$ \times
(\lambda^{(-)}_{m_1} (j_1 j_2))^{-1}
\lambda^{(+)}_{m_2} (j_1 j_2)  \lambda^{(-)}_{\ell_1} (j_1 j_2)
(\lambda^{(-)}_{\ell_2} (j_2 j_2))^{-1}$ \\
\indent \indent \indent\indent\indent\indent\indent\indent $\times a_{(0)(q)}
\left[ \matrix{ j_2 j_2 \cr j_1 j_1
\cr j_2 j_2} \right] a_{(p)(q)} \left[ \matrix{ j_2 j_1 \cr j_2 j_2
\cr j_1 j_2} \right] a_{(p)(n)} \left[ \matrix{ j_2 j_1 \cr j_2 j_2
\cr j_2 j_1} \right]$\\
\indent\indent\indent\indent\indent\indent\indent\indent $ \times  a_{(m)(n)}
\left[ \matrix{ j_2 j_2 \cr j_2 j_1
\cr j_2 j_1} \right] a_{(m)(\ell)} \left[ \matrix{ j_2 j_1 \cr j_2 j_2
\cr j_1 j_2} \right] a_{(0)(\ell)} \left[ \matrix{ j_2 j_2 \cr j_1 j_1
\cr j_2 j_2} \right]$

\vspace{.5cm}

\noindent $7_6 : \qquad [2j_1+1] [2j_2+1]^2 \sum (\lambda^{(-)}_{q_1} (j_1 j_2)
\lambda^{(-)}_{q_2} (j_2 j_2))^{-1} \lambda^{(-)}_{p_2} (j_1 j_2)$ \\
\indent\indent\indent\indent\indent\indent\indent\indent $ \times
\lambda^{(+)}_{n_1} (j_2 j_2) \lambda^{(-)}_{m_1} (j_1 j_2)
(\lambda^{(-)}_{\ell_1} (j_1 j_2) \lambda^{(-)}_{\ell_2} (j_2 j_2))^{-1}$ \\
\indent \indent \indent\indent\indent\indent\indent\indent $\times a_{(0)(q)}
\left[ \matrix{ j_1 j_1 \cr j_2 j_2
\cr j_2 j_2} \right] a_{(p)(q)} \left[ \matrix{ j_1 j_2 \cr j_1 j_2
\cr j_2 j_2} \right] a_{(p)(n)} \left[ \matrix{ j_1 j_2 \cr j_2 j_1
\cr j_2 j_2} \right]$ \\
\indent\indent\indent\indent\indent\indent\indent\indent $\times a_{(m)(n)}
\left[ \matrix{ j_1 j_2 \cr j_2 j_1
\cr j_2 j_2} \right] a_{(m)(\ell)} \left[ \matrix{ j_1 j_2 \cr j_1 j_2
\cr j_2 j_2} \right] a_{(0)(\ell)} \left[ \matrix{ j_1 j_1 \cr j_2 j_2
\cr j_2 j_2} \right]$

\vspace{.5cm}

\noindent $7_7 : \qquad [2j_1+1] [2j_2+1]^2 \sum (\lambda^{(-)}_{n_1} (j_1
j_2))^{-1}
(\lambda^{(-)}_{n_2} (j_2 j_2)) (\lambda^{(-)}_{m_1} (j_1 j_2))^{-2}$ \\
\indent\indent\indent\indent\indent\indent\indent\indent $\times
\lambda^{(+)}_{m_2} (j_2 j_2) (\lambda^{(-)}_{\ell_1} (j_1
j_2))^{-1} \lambda^{(-)}_{\ell_2} (j_2 j_2)$ \\
\indent \indent \indent\indent\indent\indent\indent\indent $\times a_{(0)(n)}
\left[ \matrix{ j_1 j_1 \cr j_2 j_2
\cr j_2 j_2} \right] a_{(m)(n)} \left[ \matrix{ j_1 j_2 \cr j_1 j_2
\cr j_2 j_2} \right] a_{(m)(\ell)} \left[ \matrix{ j_1 j_2 \cr j_1 j_2
\cr j_2 j_2} \right] a_{(0)(\ell)} \left[ \matrix{ j_1 j_1 \cr j_2 j_2
\cr j_2 j_2} \right]$

\vspace{.5cm}

\noindent $7_8 : \qquad [2j_1+1] [2j_2+1]^2 \sum \lambda^{(-)}_{q_1} (j_1 j_2)
(\lambda^{(-)}_{q_2} (j_1 j_2))^{-1} (\lambda^{(-)}_{p_1} (j_2 j_2)
\lambda^{(+)}_{p_2} (j_1 j_2))^{-1}
$ \\
\indent \indent \indent\indent\indent\indent\indent\indent $\times (\lambda
^{(+)}_{n_2}(j_2 j_2))^{-1}(\lambda^{(-)}_{m_1} (j_2
j_2))^{-1} \lambda^{(+)}_{m_2} (j_1 j_2)\lambda^{(-)}_{\ell_1} (j_1
j_2) (\lambda^{(-)}_{\ell_2} (j_1 j_2))^{-1}$ \\
\indent \indent \indent\indent\indent\indent\indent\indent $\times a_{(0)(q)}
\left[ \matrix{ j_2 j_2 \cr j_1 j_1
\cr j_2 j_2} \right] a_{(p)(q)} \left[ \matrix{ j_2 j_1 \cr j_2 j_2
\cr j_1 j_2} \right] a_{(p)(n)} \left[ \matrix{ j_2 j_1 \cr j_2 j_2
\cr j_2 j_1} \right]$ \\
\indent\indent\indent\indent\indent\indent\indent\indent $\times  a_{(m)(n)}
\left[ \matrix{ j_2 j_1 \cr j_2 j_2
\cr j_2 j_1} \right] a_{(m)(\ell)} \left[ \matrix{ j_2 j_1 \cr j_2 j_2
\cr j_1 j_2} \right] a_{(0)(\ell)} \left[ \matrix{ j_2 j_2 \cr j_1 j_1
\cr j_2 j_2} \right]$

\vspace{.5cm}

Again identity (A.13) has been used in above calculations. A corresponding
useful identity for six-point duality matrices is :
$$
\sum_{(q)} (\lambda_{q_1} (\hat{j}_1^\ast \hat{j}_2))^r (\lambda_{q_2}
(\hat{j}_2^\ast \hat{j}_3))^s  a_{(0)(q)} \left[ \matrix{ j_1 j_1 \cr j_2 j_2
\cr j_3 j_3} \right] a_{(p)(q)} \left[ \matrix{ j_1 j_2 \cr j_1 j_3
\cr j_2 j_3} \right] \hspace{4cm}
$$
$$
\hspace{3cm}= (-)^{2min(j_1,j_2) + 2min(j_2,j_3)} (\lambda_{p_{0}} (\hat{j}_1
\hat{j}_2))^{-r} (\lambda_{p_2} (\hat{j}_2
\hat{j}_3))^{-s}  a_{(p)(0)} \left[ \matrix{ j_2 j_1 \cr j_1 j_3
\cr j_3 j_2} \right]
$$

\noindent where $r = \pm1, s = \pm1 $.

For mirror reflected links, the invaraints are obtained by conjugation.

Notice for $q = 1$ (that is, $k \rightarrow \infty$ ), every one of these
invariants
reduces to the product of the ordinary dimensions of the representations placed
on the two component knots, $V_{j_1 j_2} ( q = 1) \ = \ (2j_1+1)(2j_2+1)$.

\newpage

\noindent {\bf Figure Captions}

\begin{description}
\item[Fig.1.] Two ways of combining four spins into singlets
\item[Fig.2.] Two ways of combining 2m spins into singlets
\item[Fig.3.] Duality transformation of 6-point correlators
\item[Fig.4.] Duality transformation of 8-point correlators
\item[Fig.5.] Examples of coloured-oriented braids
\item[Fig.6.] An oriented-coloured n-braid ${ \cal B} \pmatrix{ \hat{j}_1 &
\hat{j}_2 &
\ldots & \hat{j}_n \cr \hat{\ell}_1 & \hat{\ell}_2 & \ldots & \hat{\ell}_n}$
\item[Fig.7.] Identity braids $I \pmatrix{ \hat{j}_1 & \hat{j}_2 &
\ldots & \hat{j}_n \cr \hat{j}_1^\ast & \hat{j}_2^\ast & \ldots &
\hat{j}_n^\ast}$ and braid generators \\
 ~~~~~~${ B_\ell} \pmatrix{ \hat{j}_1 & \ldots &
\hat{j}_\ell & \hat{j}_{\ell+1} & \ldots & \hat{j}_n \cr \hat{j}_1^\ast &
\ldots & \hat{j}_{\ell+1}^\ast & \hat{j}_\ell^\ast & \ldots & \hat{j}_n^\ast}$
\item[Fig.8.] Relations among braid generators
\item[Fig.9.] Platting of a coloured-oriented braid
\item[Fig.10.] Closure of a coloured-oriented braid
\item[Fig.11.] An indentity braid in a three manifold with boundaries
$\sum^{(1)}$ and ~~~~~~$\sum^{(2)}$, each an $S^2$
\item[Fig.12.] Braid generators introduce (a) right-handed half-twists in \\
{}~~~~~~parallely oriented strands and (b) left-handed half-twists in
anti-parallel
strands
\item[Fig.13.] A three-manifold containing an arbitrary coloured oriented
$2m$-baid \\
 ~~~~${\cal B}$
\item[Fig.14.] Three-balls containing $m$ Wilson lines
\item[Fig.15.] $m$-braid$ {{\cal B}_m}$ extended to $2m$-braid ${\hat {\cal
B}}_m$
\item[Fig.16.] (a) Borromean rings and (b) a plat representation for it
\item[Fig.17.] Consistency conditions on ${ \lambda^{(\pm)}_j} (j_1 j_2)$ and
${a_{j\ell}} \left[ \matrix{j_1 & j_2 \cr j_3 & j_4 } \right]$

\item[Fig.18.] A cycle of five duality transformations
\item[Fig.19.] Plat representations for knots upto eight crossing number
\item[Fig.20.] Plat representation for two-component links upto seven
crossing\\
{}~~~~number.
\end{description}

\end{document}